\documentclass[reprint,pre,aps]{revtex4-1}
\usepackage{graphicx}
\usepackage{amsfonts}
\usepackage{amsmath}
\usepackage{psfrag}
\usepackage{natbib}
\usepackage{color}
\usepackage{mathrsfs}
\usepackage{bm}
\usepackage{color}
\usepackage{subfigure}
\usepackage{gensymb}

\usepackage[T1]{fontenc}
\usepackage[english]{babel}

\newcommand{\beq}{\begin{equation}}
\newcommand{\eeq}{  \end{equation}}
\newcommand{\beqa}{\begin{eqnarray}}
\newcommand{\eeqa}{  \end{eqnarray}}

\newcommand\bn{\bm{n}}
\newcommand\bv{\bm{v}}
\newcommand\bh{\bm{h}}
\newcommand\bt{\bm{t}}
\newcommand\bl{\bm{\ell}}

\definecolor{red}{rgb}{1,0,0}
\definecolor{blue}{rgb}{0,0,1}
  
\newcommand{\blue}{\color{blue}}
\graphicspath{{./Figs/}}

\usepackage{textcomp}
\usepackage[sb]{libertine}
\usepackage[varqu,varl]{zi4}
\usepackage[libertine,bigdelims,vvarbb]{newtxmath} 
\usepackage[cal=boondoxo]{mathalfa} 
\usepackage[supstfm=libertinesups,%
	supscaled=1.2,%
raised=-.13em]{superiors}

\def\strutdepth{\dp\strutbox}
\def\nw#1{\strut\vadjust{\kern-\strutdepth\vtop to0pt{\vss\hbox to\hsize
{\hskip\hsize\hskip5pt$\leftarrow$\hss\strut}}}{\blue \em #1}}

\begin{document}

 \title{Tree crowns grow into self-similar shapes controlled by gravity and light sensing} 

\author{Laurent Duchemin}
\affiliation{Aix Marseille Univ, CNRS, Centrale Marseille, IRPHE, Marseille, France}
\author{Christophe Eloy}
\affiliation{Aix Marseille Univ, CNRS, Centrale Marseille, IRPHE, Marseille, France}
\author{Eric Badel}
\affiliation{UCA, INRA, UMR PIAF, F-63000 Clermont-Ferrand, France}
\author{Bruno Moulia}
\affiliation{UCA, INRA, UMR PIAF, F-63000 Clermont-Ferrand, France}

\date{\today}

\begin{abstract} 
Plants have developed different tropisms: in particular, they re-orient the growth of their branches towards light (phototropism) or upwards (gravitropism). How these tropisms affect the shape of a tree crown remains unanswered. We address this question by developing a propagating front model of tree growth.
This model being length-free, it leads to self-similar solutions, independent of the initial conditions, at long time. Varying the intensities of each tropism, different self-similar shapes emerge, including singular ones. Interestingly, these shapes bear similarities with existing tree species. It is concluded that the core of specific crown shapes in trees relies on the balance between tropisms. 
\end{abstract}

\maketitle

\section{Introduction}

In many tree species the outer shape of the crown is a criterion for species identification, especially when trees have grown in isolation. 
The Mediterranean cypress has a  distinctive elongated shape, while birch and oaks show more spherical shapes, and the Norway spruce has a conical profile.
The growth process that leads to these different silhouettes is not well understood. It is associated with  genetically fixed species-dependent factors like the branching angles, the phyllotaxy, and more generally the developmental characteristics of the branching architecture known as the ``architectural model'' of the species \cite{Barthelemy2007}. But crown shapes are also affected by the environmental growth conditions (in a genetic-dependent manner): trees grow differently if they are isolated or in a forest \cite{Cournede2008}, if subject to climatic stresses like wind or snow cover \cite{Davis1980}. So far, the few studies about crown shaping have investigated the hypothesis that it may result from an interplay between a genetically-defined architectural development and branch growth or shedding linked to light competition \cite{Mech1996}.
 
However tropisms in the outer growing branches can be viewed as another possible candidate for the control of crown shape. Tropisms are defined as the re-orientation of a growing branch following a vectorial cue from the environment. Two tropisms are shared by most plants: phototropism is the process that leads to growth in the direction of light, while gravitropism is driven by the direction of earth gravity. Genetic differences in  tropic sensitivities have been reported \cite{sierra1997, Bastien2013}. The study of these tropisms is still an active field today, ranging from shoot scale \cite{Bastien2013,Chauvet2016,Riviere2017, Chelakkot20170001}, to the  molecular networks that regulate them \cite{Goyal2013}. Yet, a question remains largely open: how do these tropisms and their balance influence the shape of a tree crown?  

In this work we investigate the simple hypothesis that the two tropisms of the peripheral primary shoots at the tip of the branches can be major players in driving the shaping of the crown, accounting for the interplay between genetic control and responses to the local environment. To do so we developed a simple model of the crown growth. In the literature, popular growth models generally consider branches and leaves as the elementary bricks of a numerical simulation \cite{Perttunen1998,Allen2005,Barczi2008,Palubicki2009,Guo2011}. However these models usually involve a large number of empirical parameters, and thus remain too complex to study the specific roles of phototropism and gravitropism in selecting the tree shapes. Recently, Beyer\cite{Beyer2014} proposed a simpler and more parsimonious model that considers the whole tree as a continuous medium. This approach seems very promising, but has not yet been fully exploited to address the combined role of phototropism and gravitropism. 
In this Letter, we propose a new idea: considering the growth of a tree crown as a continuous front propagation process. Despite some similarities with other front propagation processes, such as crystal growth \cite{brower1983geometrical}, premixed combustion \cite{clavin2016combustion}, or Saffman--Taylor fingering \cite{Pelce2012a}, tree growth has some distinctive features. 

In particular, since growth involves photosynthesis and is controlled through light sensing (photomorphogenesis), our model includes both photosensitivity (the foliage receiving more light will grow faster), phototropism (growth is preferentially oriented in the direction of light), {and also gravitropism (growth is preferentially oriented against gravity).}
As we shall see below, the proposed model requires only two parameters to describe the shape of the crown (irrespective of its size): the intensities of phototropic and gravitropic growing responses. With this model, we intend to address the following questions: What is the family of shapes generated when these two parameters vary? Do these shapes converge towards steady solutions? How these steady solutions can be described analytically? And finally do these solutions fit the major classes of crown shapes observed in undisturbed isolated trees in nature?

\section{Crown growth model}

\begin{figure}[h!]
\begin{center}
\includegraphics[]{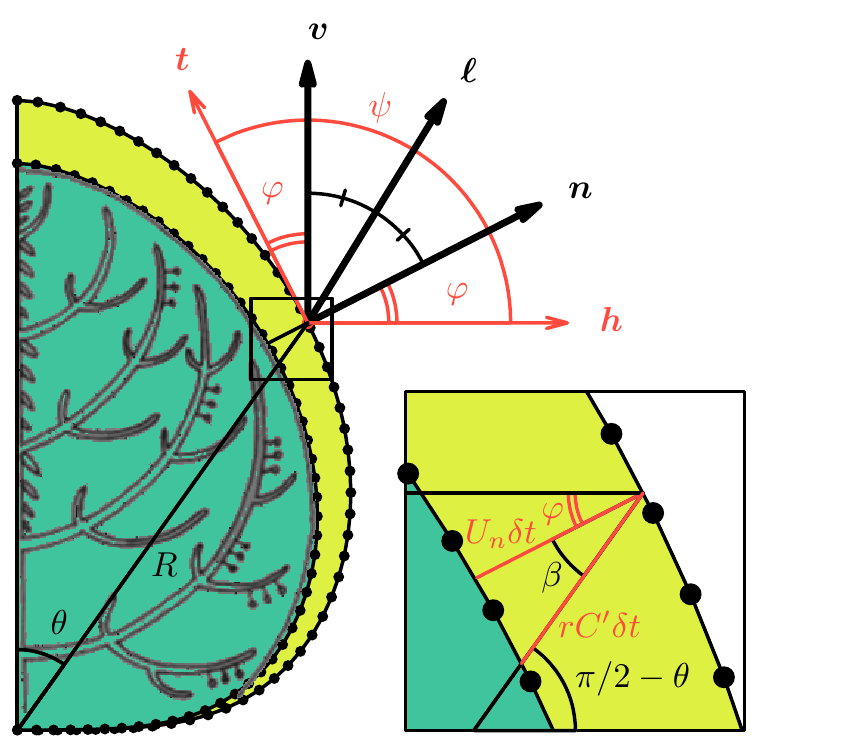}
\end{center}
\caption{Sketch of a growing tree crown. $\bh$ and $\bv$ are unit horizontal and vertical vectors respectively, $\bn$ and $\bt$ are the unit vectors   respectively  normal and tangent  to the front and the unit vector $\bl$ points towards the mean direction of light, which is the first bisector of $\widehat{(\bh,\bt)}$. The angle $\psi = \varphi+\pi/2$ represents the local amount of sunlight intercepted for this axisymmetric shape.
The inset shows a zoom around the front to highlight the conditions for self-similarity of the growing shape.}
\label{sketch}
\end{figure}

We first assume that the average sunlight is uniformly distributed in the upper hemisphere, but we  
shall come back to this strong assumption in the discussion section.
When the crown envelope is convex and axisymmetric, the amount of light received daily at each point of the front is thus assumed to be proportional to $\psi$, the angle  between the local tangent and the horizontal (Fig.~\ref{sketch}).
At each point on the front, $\bn$ represents the outward normal unit vector, $\bl$ is the unit vector that represents the average direction of light, $\bv$ is the upward vertical  unit vector, and $\kappa$ is the in-plane curvature, positive when the surface is locally concave (Fig.~\ref{sketch}).
According to this nomenclature, the velocity of the front is written as
\beq
\bm{U} = \psi \frac{\bm{n} + \alpha_g \bm{v} + \alpha_p \bm{\ell}}{\left| \bm{n} + \alpha_g \bm{v} + \alpha_p \bm{\ell} \right|} + \gamma \kappa \bm{n},
\label{velocity}
\eeq
where $\alpha_g$ and $\alpha_p$ are the intensities of the crown gravitropic and phototropic responses respectively, and $\gamma$ is similar to a ``surface tension'', but acting on the growth velocity. 

Our model is based on two basic assumptions: {\bf [A1]} the crown foliage is dense enough all across the crown so that it may be described as a continuous medium with a continuous boundary line (i.e., a growth front); {\bf [A2]} all the growing apices sitting on the front have similar dependency on light and gravity.
Assumption {\bf [A1]} requires that the development of the tree crown is sufficient and the crown displays a steady characteristic shape, a condition that is usually observed in most species as soon as they have reached the ``architectural unit'' (AU) stage, i.e. that they have differentiated all their branch types \cite{edelin1996}.
{\bf [A2]} requires that ``rogue branches'' do not occur. It is well known that traumatic reiteration resulting from branch accidental breakage or from large herbivory damages produces re-juvenilised branches that behave differently than regular branches (e.g., suckers). 
These traumatic reiterations should be distinguished from regular reiterations involved in the process of crown morphogenesis (a process known as branch metamorphosis \cite{Barthelemy2007, edelin1996}), as these reiterated branches still complie with a regular behaviour and the crown remains dense. 

The first term on the right-hand-side of Eq.~\eqref{velocity} includes two important biological mechanisms.  
First, the front velocity is assumed to be proportional to $\psi$, the light intercepted, to account for photosensitive growth. 
Second, the gravitropism and the phototropism are modelled through a re-orientation of the growth in a direction computed as the weighted average of the three unit vectors $\bn$, $\bv$, and $\bl$. {This modelling of tropisms is in accordance with the experimental observations made at the level of a single shoot, and with the ArC model of shoot gravi-photo-tropism, that was assessed against experiments~\cite{Bastien2015}.}
Additionally, it has been demonstrated experimentally that the sensitivity to the gravity is not dependent on the angle made by the shoot versus gravity. 
The mechanism subtending this behaviour lies in the functioning of specialized cells called statocytes (see \cite{Bastien2013, Dumais2013, Chauvet2016, pouliquen2017} for more explanations on the physics and biology behind this mechanism). For phototropism, the equation and the definition of phototropic sensitivity are mainly similar (see \cite{Bastien2015} and references therein).

 
The last term in Eq.~\eqref{velocity} stabilises the front dynamics and smoothes out the front shape by damping the velocity fluctuations, and hence has a similar role as surface tension in Saffman--Taylor dynamics for example. There is obviously no surface tension acting at the boundary of the tree crown. However recent ecophysiological investigations have revealed that two biological mechanisms are acting  as to reduce the differences in growth velocities of neighbouring shoots, and hence resulting in a flattening tendency of the canopy boundary, which is qualitatively similar to a surface tension. The first mechanism is due to lateral sensing of the  spectral signature of the light reflected by neighbouring plants, through the phytochrome pigment \cite{pierik2013}. 
This sensing results in a photomorphogenetic synchronisation of the growth in length of the neighbouring stems, keeping the top of the canopy flat.  The second mechanism involves the sensing of wind-induced strains~\cite{moulia2015}. Whenever a shoot overreach its neighbours, it is not sheltered by the canopy anymore and its growth speed is reduced until the front is flat again. It has been shown nicely in herbaceous populations of shoots that these two mechanisms are responsible for the flattening tendency of canopy tops, as dramatically illustrated by the flatness of the top surface of crops such as wheat or corn~\cite{nagashima2012}. 
On a practical point of view, when simulating the growth of a crown, we rescale its shape at each time-step to keep its axi-symmetric volume $V$ constant.
As explained in the next section, this is equivalent to a coefficient $\gamma$ growing in time in Eq.~\eqref{velocity}. Although this rescaling may seem unphysiological, we perform it for two reasons: it ensures that the code is stable at long times, and it allows to reach self-similarity (when surface tension is rescaled, Eq.~\eqref{velocity} has no length scale and self-similar solutions emerge).

Finally it may be noted that in our simplified model, the curvature $\kappa$ is calculated in a vertical plane, and not as the three-dimensional mean curvature of the axisymmetric interface. The latter refinement would be possible, but does not affect the main features of the solutions. 

\section{Numerical solutions}

\begin{figure}[t!]
\begin{center}
\includegraphics[]{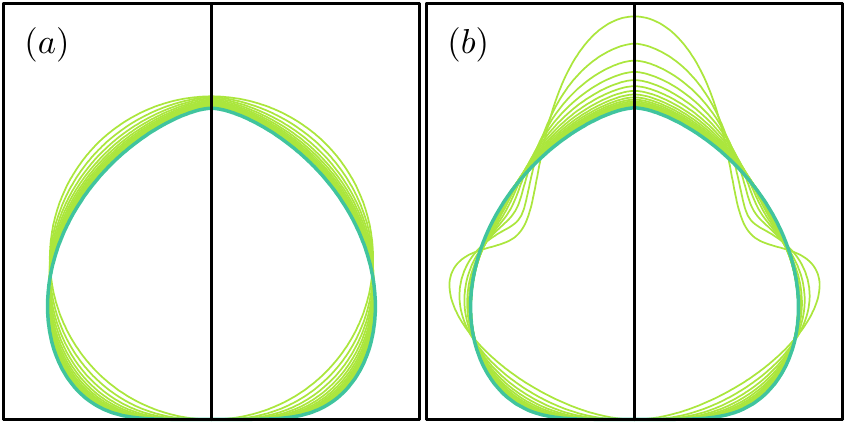}
\end{center}
\caption{Superimposed views of a tree crown growth for two different initial conditions. In (a)  the initial shape is a circle and in (b) a circle with a significant $\mathcal{O}(1)$ perturbation. Parameters are: $\alpha_g=\alpha_p=0$ and $\gamma=0.01$. The light green curves correspond to different instants. The dark green curves correspond to the self-similar solution at large time.
The shape is rescaled at each time-step in order to keep its volume constant equal to one.
} 
\label{evolution}
\end{figure}

\begin{figure*}[t!]
\begin{center}
\includegraphics[]{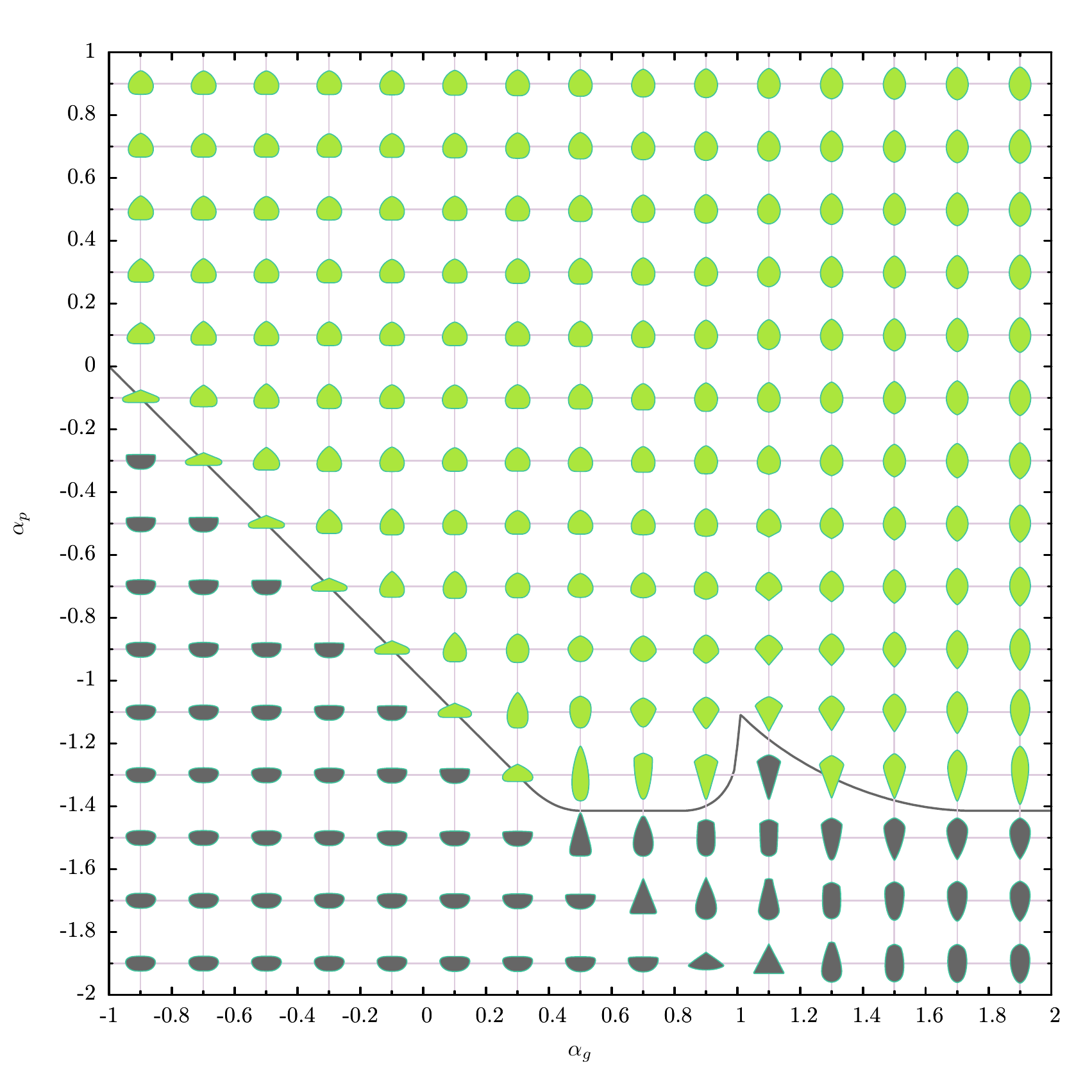}
\end{center}
\caption{A set of crown shapes obtained from a numerical computation, varying $\alpha_g$ and $\alpha_p$, and with $\gamma=0.01$. Green shapes have converged towards a self-similar-one, whereas grey shapes have not converged.
	{For each shape, the parameter values correspond to the grid intersection.}
}
\label{graph}
\end{figure*}

Equation~\eqref{velocity} is solved numerically by placing $N=200$ marker points on the front. 
The points are then advected along the normal to the front. 
Furthermore, at each time-step, the coordinates of the front are rescaled by the volume to the power $1/3$, such that the crown volume remains constant ($V=1$). This rescaling makes the problem scale-invariant and allows the ``surface tension'' to remain of the same order during the growth process. As a consequence, the scaling of the true curvature in the physical plane increases like $V^{1/3}$. Since the ``surface tension'' coefficient in the numerical space $\gamma^*$ is kept constant, it means that~:
\begin{equation}
\gamma = \gamma^* V^{1/3}.
	\label{gamma_star}
\end{equation}
Since $V=1$, we can drop the star and use $\gamma$ in place of the constant $\gamma^*$. 

The surface points are advected in time using an explicit second-order scheme  \cite{Duchemin2014}, with a time-step $\delta t \sim \delta s^2/\gamma $, where $\delta s$ is the typical distance between two successive marker points, chosen to ensure the stability of this explicit scheme. 
The markers are redistributed on the front at each time-step, in order to conserve a regular spacing. Note that the ``surface tension'' becomes significant when $\gamma \kappa \sim \mathcal{O}(1)$, as seen in Eq.~\eqref{velocity}.

Figure~\ref{evolution} shows two examples of such a computation for two distinct initial conditions. In this example, there is no phototropism and no gravitropism ($\alpha_g=\alpha_p=0$) and surface tension is set to $\gamma=0.01$, small compared to the global length scale: $\gamma \ll 1$. 

We first observe that, although initial conditions significantly differ, the successive shapes converge towards a unique self-similar shape (Fig.~\ref{evolution}a). This self-similarity was expected because the problem has been made scale-invariant.
To further test this idea, we  performed extensive numerical simulations with different parameter sets ($\alpha_g$, $\alpha_p$), at low surface tension ($\gamma=0.01$). The results are presented in figure \ref{graph}: green shapes correspond to computations that have converged towards self-similar shapes, meaning that a typical distance between two successive rescaled shapes (Fréchet distance) has reached $10^{-6}$, whereas grey shapes correspond to non-convergent computations 
(in these cases found in the lower-left part of the diagram , the shape flattens out indefinitely and convergence is never reached). 
The grey curve which segregates converged from non-converged shapes will be discussed in section \ref{s_comparison}.Figure \ref{graph} displays a large diversity of shapes as a function of the parameter sets ($\alpha_g$, $\alpha_p$), showing that the (genetic) variation in the sensitivities to phototropism and gravitropism can indeed produce different crown shapes in our model.

Moreover, we  performed numerical simulations with decreasing surface tension $\gamma$ (Fig.~\ref{tension}). The results showed that all the final self-similar shapes converge towards a universal shape in the limit $\gamma \to 0$.
 In the next section, we focus on this limit and show that this universal shape can be derived analytically, providing  insight into the control of the steady-state crown shape, when it exists, by the parameter sets ($\alpha_g$, $\alpha_p$).

\section{Analytical solutions}
In the limit of vanishing surface tension (i.e. $\gamma \to 0$), Eq.~\eqref{velocity} becomes scale invariant. We look for a self-similar solution of this equation by imposing that the front is described by the homothetic surface:
\beq
R(\theta,\phi,t)= r(\theta) C(t),
\eeq 
in spherical coordinates (Fig.~\ref{sketch}).

Between two successive times, $t$ and $t+\delta t$, the front has been increased by $r(\theta)C'(t) \delta t$ in the radial direction.
This condition can be expressed as
\beq
r(\theta) C'(t) \delta t \cos\beta = \bm{U}\cdot\bm{n}\, \delta t
= \psi B(\varphi) \,\delta t,
\label{homothety}
\eeq
where $\psi = \varphi + \pi/2$, $\beta = \pi/2 - \theta - \varphi$ and 
\beq
B(\varphi)
= \frac{\bm{n} + \alpha_g \bm{v} + \alpha_p \bm{\ell}}{\left| \bm{n} + \alpha_g \bm{v} + \alpha_p \bm{\ell} \right|} \cdot \bm{n}.
\eeq

\begin{figure}[b!]
\begin{center}
\includegraphics[width=0.95\linewidth]{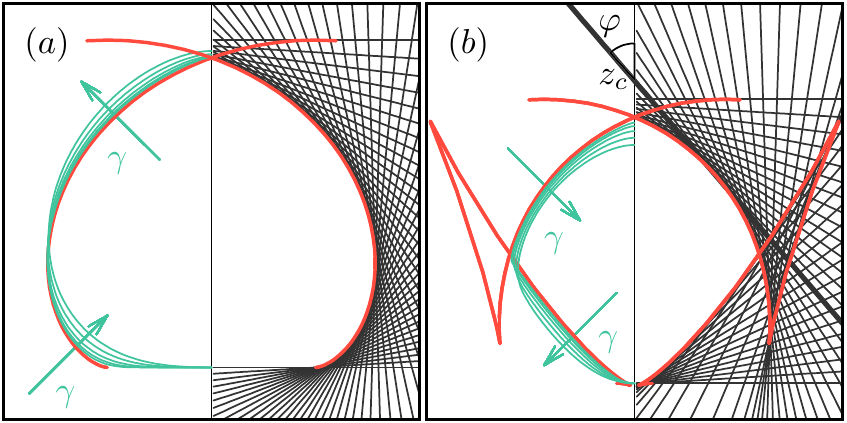}
\end{center}
\caption{Attracting shapes obtained for different values of $\gamma$: $\gamma=0.04, 0.02, 0.01, 0.005$ and $0.0025$, with $\alpha_g=\alpha_p=0$ (a) and $\alpha_g=0.8, \alpha_p=-1$ (b). The envelope of the grey lines given by equation (\ref{envelope}) is the expected self-similar shape. When $\gamma \to 0$, the time-evolving shapes converge towards this self-similar shape.} 
\label{tension}
\end{figure}
\begin{figure*}[t!]
\begin{center}
\includegraphics[]{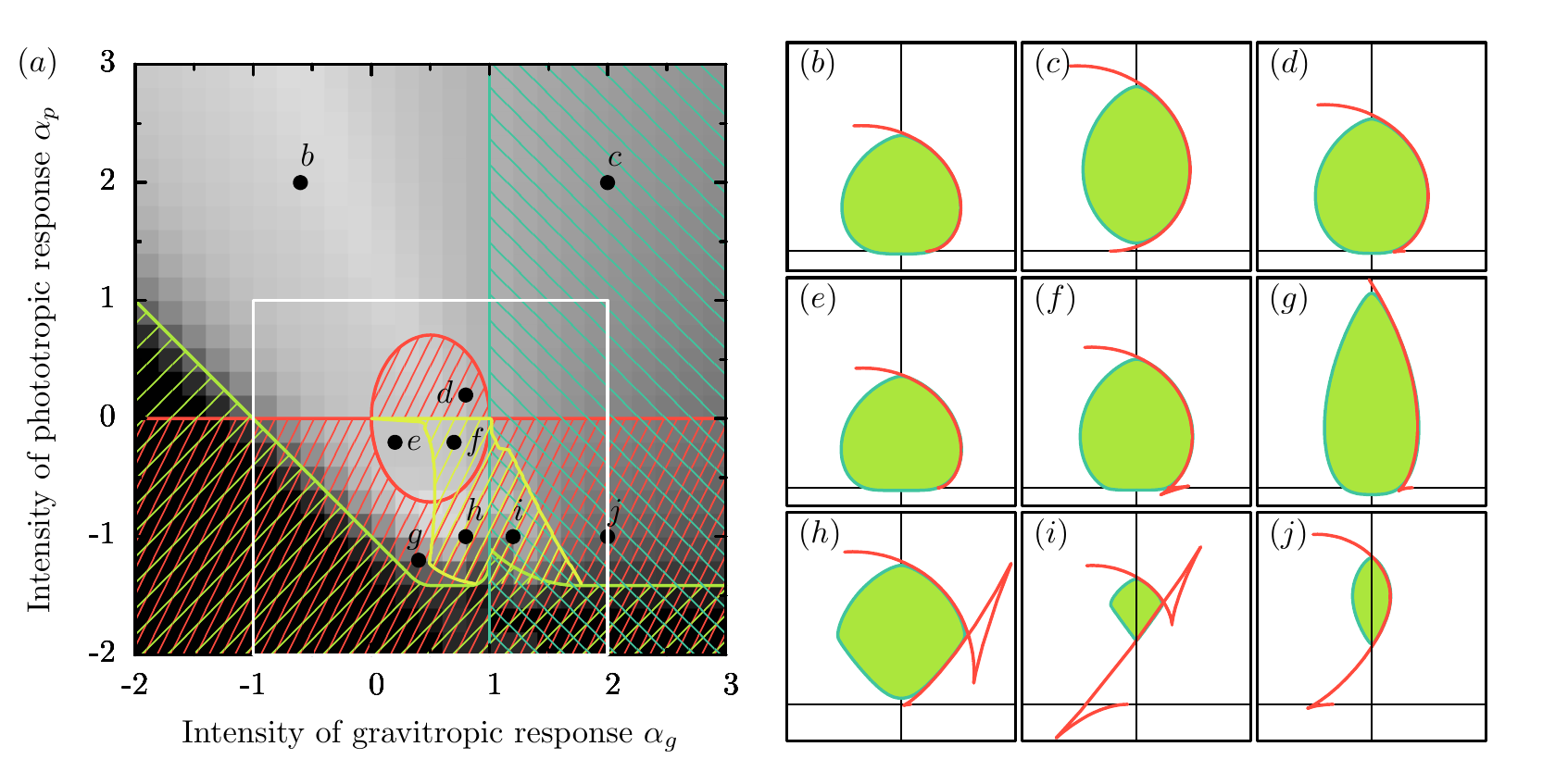}
\end{center}
\caption{(a) Phase diagram in the $\alpha_g$--$\alpha_p$ plane. The grey level is proportional to the number of time-steps necessary for convergence towards a self-similar shape (black corresponds to no convergence). 
The hatchings correspond to different features of the self-similar shapes: light green for no self-similar solutions, red when a cusp is present at the bottom, yellow for a loop, and dark green for a pointed bottom (more details in the main text).  The white square corresponds to figure \ref{graph}.
(b--j) Characteristic shapes found numerically for $\gamma=0.01$ (green curve and fill) and analytically in the limit of vanishing surface tension with Eqs.~(\ref{eq:x}--\ref{eq:z}) (red curve). The thin black lines correspond to the vertical symmetry axis $(Oz)$ and to the horizontal plane $z=0$. Shapes (i) and (j) have been rescaled by a factor of $2$ in both directions, whereas all others shapes have the same volume.} 
\label{diagram}
\end{figure*}

To solve this problem, we first note that cones are trivial self-similar solutions of Eq.~\eqref{homothety}, with $\varphi$ the constant angle between the cone surface and the vertical direction (Fig.~\ref{sketch}). 
Moreover, Eq.~\eqref{homothety} implies that $C'(t)$ does not depend on time. 
The cones having the same velocity $C'$ (that we choose equal to $1$ without loss of generality) intersect the vertical axis at the elevation
\beq
z_c(\varphi) = r(0) = {(\varphi+\pi/2)B(\varphi)}/{\sin\varphi}. 
\label{envelope}
\eeq

Figure~\ref{tension} shows the cones described by Eq.~\eqref{envelope} for two different sets of parameters. The universal shape we found numerically in the limit of vanishing surface tension appears as the inside envelope of all those cones. Interestingly, this inside envelope can be obtained analytically.

For convenience, we switch to cartesian coordinates. The cones of Eq.~\eqref{envelope} intersect the plane $y=0$ along the straight line described by the equation: $z = z_c(\varphi) - x / \tan \varphi$. Consequently, two cones corresponding to angles $\varphi$ and $\varphi + \delta\varphi$  will intersect at the point \begin{eqnarray}
x (\varphi)& = & -z'_c(\varphi) \sin^2\varphi, \label{eq:x}\\
z (\varphi)& = & z_c(\varphi) + z'_c(\varphi) \sin\varphi \cos\varphi.  \label{eq:z}
\end{eqnarray}
When $-\pi/2\leq\varphi\leq\pi/2$, Eqs.~(\ref{eq:x}--\ref{eq:z}) describe the cone envelope as a parametric curve (red curves in Fig.~\ref{tension}). Because this parametric curve is everywhere tangent to a cone with the same time evolution $C(t)$, it is also a self-similar solution of Eq.~\eqref{velocity} for $\gamma=0$. This approach is similar to the Wulff construction used in the physics of crystal growth \cite{Einstein2015,Pimpinelli1998}.

\begin{figure}[h!]
\centering
\begin{tabular}[c]{cc}
\includegraphics[]{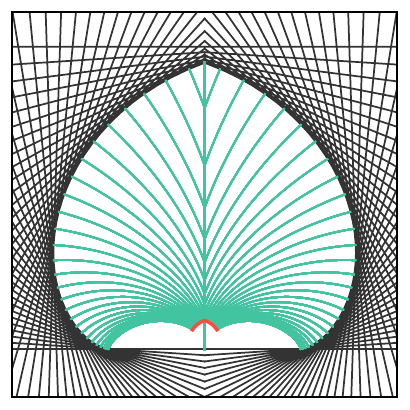} & \includegraphics[]{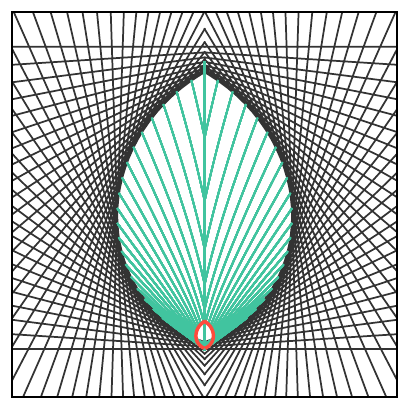}
\end{tabular}
\caption{Trajectories of points along the front, for the cases $\alpha_g=\alpha_p=0$ (left) and $\alpha_g=\alpha_p=2$ (right). Starting from the self-similar profile corresponding to $t=1$, $31$ points are advected backward in time using Eq.~\eqref{velocity} until $t=0.1$.}
\label{trajectories}
\end{figure}

Another interesting feature of these self-similar shapes arises from following the trajectories of surface points backward in time. These trajectories can be thought of as traces of the underlying structure, {\it i.e.} the tree branches. Figure~\ref{trajectories} shows these trajectories for two cases: $\alpha_g=\alpha_p=0$ and $\alpha_g=\alpha_p=2$. 
These trajectories are numerically calculated by advecting backward in time $31$ points along the vector $-\bm{U}$. At each time step, the points --lying on a smaller self-similar shape-- are associated to an updated angle $\psi$ using cubic splines.
For $\alpha_g=\alpha_p=0$, there is a region without branches close to the bottom. This region is a consequence of the flat bottom, where $\psi=0$ and the front velocity is zero. Note that the curvature of the ``branches'' is only due to the variation of $\psi$ and $\bm{\ell}$ along these trajectories, and not to the bending under self-weight of the branches or any other global re-orientating mechanisms such as those related to reaction wood\cite{moulia2009}, which are not taken into account in our model. 

We will discuss below why numerical simulations converge towards the parametric solution given by Eqs.~(\ref{eq:x}--\ref{eq:z}). Meanwhile, we describe how the respective intensities of gravitropic and phototropic responses modify the crown propagation and the final shape.

\section{Comparison between numerical and analytical solutions}
\label{s_comparison}

We note that these attracting shapes do not exist for any choice of the parameters $\alpha_g$ and $\alpha_p$, as already observed in figure \ref{graph}. Figure \ref{diagram}a shows with  grey levels how fast the numerical solution converges towards a self-similar solution. {It also shows that in the lower part of this diagram, there is no convergence.}
Equation~\eqref{envelope} indicates the cases when self-similar shapes exist. 
The convergence occurs when the cone with a vertical angle ($\varphi = 0$) exists (i.e. $x(0) > 0$ in Eq.~\ref{eq:x}). This condition simplifies to:
\beq
\alpha_p > -\sqrt{2}
\label{boundary1}
\eeq
Moreover, the self-similar shapes cease to exist when their top (given by $z'_c(\varphi)=0$, $\varphi \in [0,\pi/2]$) reaches their bottom ($z'_c(\varphi)=0$, $\varphi \in [-\pi/2,0]$). This condition can be computed numerically and corresponds to the light green curve in figure \ref{diagram}a. 
A good approximation of this curve for $\alpha_g \in [-2,\sqrt{2}-1]$ can be obtained by writing that the self-similar shapes exist roughly when the two cones with horizontal angles ($\varphi=\pm \pi/2$) are one above the other (i.e. $z_c(\pi/2)> z_c(-\pi/2)$ in Eq.~\ref{envelope}), a condition that simplifies to:
\beq
\alpha_p + \alpha_g > -1
\label{boundary2}
\eeq
These existence conditions correspond to the light green curve in the phase diagram (Fig.~\ref{diagram}), below which there is no analytical self-similar solution. The curve is also reproduced in figure \ref{graph} and is perfectly consistent with the convergence region observed for the time-evolving shapes. 

When they exist, self-similar solutions exhibit different characteristics that correspond to different hatchings in Fig.~\ref{diagram}a. First, we note that the self-similar solutions always feature a pointed top. It happens for $z'_c(\pi/2)> 0$, with $z_c$ given by Eq.~\eqref{envelope}. This condition is equivalent to $\alpha_p + \alpha_g  > -1$; which is guaranteed by Eq.~\eqref{boundary2}. 

Second, the self-similar shape can exhibit a pointed bottom when $z'_c(-\pi/2)> 0$, or equivalently when $x(-\pi/2)< 0$. This configuration is equivalent to $\alpha_g > 1$ (dark green hatching on the right of Fig.~\ref{diagram}a). Examples of such shapes are displayed in Figs.~\ref{diagram}c,i,j. Note that, contrary to flat-bottomed shapes (Figs.~\ref{diagram}b,e,g, for instance), these shapes have a non zero growth velocity at their pointed bottom. 

Third, the bottom of the self-similar shape can present a cusp when $x'(-\pi/2)\le 0$; which is equivalent to $\alpha_p (2 \alpha_g^2 - 2\alpha_g + \alpha_p^2)\le 0$ (red hatching in Fig.~\ref{diagram}a). This cusp can clearly be seen in the self-similar solutions displayed in Figs~\ref{diagram}i,j where the value of $x$ at the bottom (for $\varphi=-\pi/2$) initially decreases. The same cusp is also perceptible in Figs~\ref{diagram}d,g,h, although it is less  pronounced. 

Finally, the self-similar shape described by the parametric curve given in Eqs.~(\ref{eq:x}--\ref{eq:z}) can exhibit a loop, as seen in Figs.~\ref{diagram}h,i (it is also perceptible in Fig.~\ref{diagram}f, but less visible). This loop corresponds to the zone with the yellow hatching in Fig.~\ref{diagram}a, whose boundary has been determined numerically. 

As seen in Figs.~\ref{tension} and \ref{diagram}b--i, the numerical self-similar solutions obtained at long time always converge, in the limit of vanishing surface tension, towards the inward envelope of the cone solutions given by the parametric curve of Eqs.~(\ref{eq:x}--\ref{eq:z}). However, when surface tension is zero, any shape constructed as an assembly of pieces of cones could be a valid self-similar solution. These solutions do not occur in the numerical simulations for two reasons: first, any connection between two pieces of cones tend to ``recess'' towards the inward envelope when the surface tension is non zero. Second, the front propagation is described as the advection of markers in the direction normal to the front, whereas a special numerical treatment of the corners would be necessary to maintain a  solution made of several assembled cones. 

\section{Comparison with observed crown shapes}
\begin{figure*}
\centering
\begin{tabular}[c]{cc}
\includegraphics[]{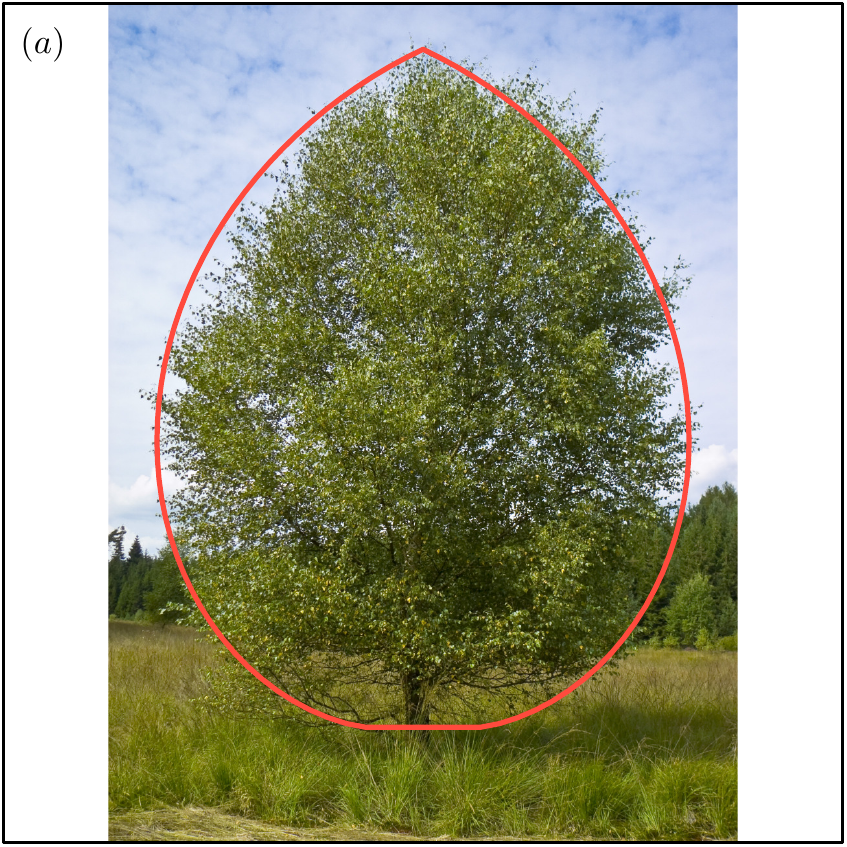} & \includegraphics[]{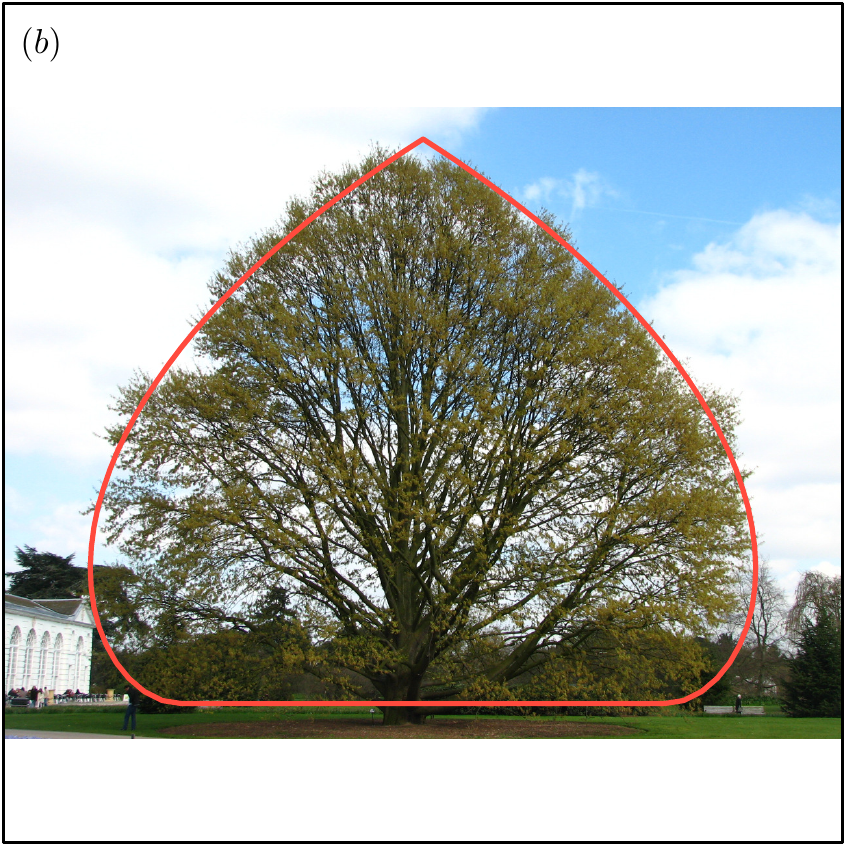} \\
\includegraphics[]{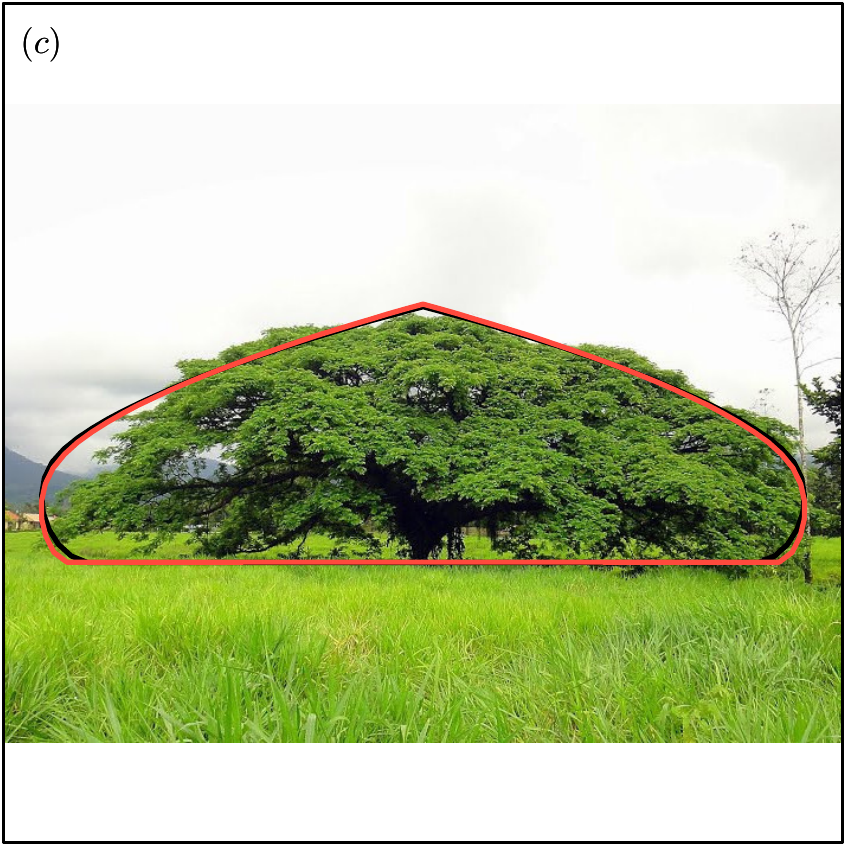} & \includegraphics[]{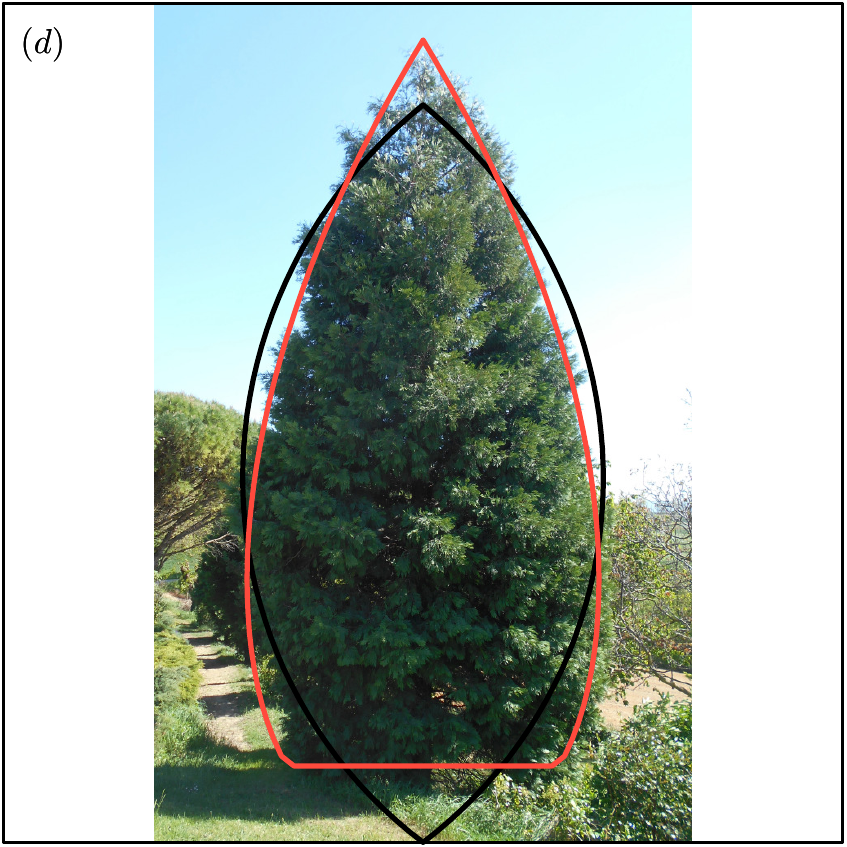}
\end{tabular}
\caption{{Comparison between real tree crowns and self-similar shapes of the model: 
	(a): \emph{Betula pubescens}, $\alpha_g=0.927, \alpha_p=0.267$, $d=0.022$;
	(b): \emph{Quercus castaneifolia}, $\alpha_g=-1.00, \alpha_p=0.353$, $d=0.019$; 
	(c): \emph{Enterolobium cyclocarpum}, $\alpha_g=-0.490, \alpha_p=-0.488$, $d=0.023$ (black curve: $\alpha_g=-1.00, \alpha_p=0.034$, $d=0.020$), 
	(d): \emph{Thuja occidentalis}, $\alpha_g=0.379, \alpha_p=-1.23$, $d=0.018$ (black curve: $\alpha_g=5.00, \alpha_p=2.69$, $d=0.079$).
	The red curve represents the best fit and the black curve the best fit with $\alpha_p>0$. 
	See Supplementary materials for comparisons with other tree species. 
}
} \label{comparison}
\end{figure*}
Care should be taken when comparing the predictions of the model with real tree crowns observed in nature. Indeed the impinging of light rays on the crown should not be intercepted by a neighbouring object (wall, rock, other trees). Moreover tree shoot should not be able to sense the reflected light from neighbouring crowns even in the absence of direct shading of the incident light.  The few studies about the distance of neighbouring sensing is limited to approximately 5~m~\cite{moulia1999}. Finally no external action such as pruning, grazing or wind-break should have occurred (at least at a scale that could affect the growth and shaping of the crown).

Another important aspect is that the external growing shoots should have a similar physiological and developmental status. Indeed some reiterations (especially traumatic ones) can result in obvious disruptions of the regularity of the crown shape. 
Therefore, only two  developmental stages were retained: (i) young trees approaching the ``architectural unit stage'' and (ii) mature trees having developed their scaffold limbs through the process of developmental metamorphosis~\cite{edelin1996,Barthelemy2007}.

We used two sources of tree crown pictures. The first source is     
an arboretum (E. Badel's park, sitting in southern France in a sub-Mediterranean climate at 44.904$\degree$N, 4.830$\degree$E) where no pruning or grazing has ever been achieved and trees were planted at large distances from each other.  
The second source results from a search on botanical websites and Wikimedia Commons (https://commons.wikimedia.org), retaining only photographs for which the species and the location were known and the criteria concerning the absence of significant neighbours or traumatic reiteration could be assessed. 

For each picture, the best fit is found using the following method. First, the right half part of the tree crown is described as a series of hand-picked points (typically about 20 points). Then, we assume that the shape is axisymmetric and we rescale and shift this shape such that its volume is equal to unity and its center of mass is at the origin. After that, we look for the parametric curve described by Eqs.~(\ref{eq:x}--\ref{eq:z}) that minimises the distance with the real tree crown, where the distance $d$ is defined as
\beq \label{eq:distance}
d^2 = \frac{1}{\pi}\int_0^{\pi} \left(R_\textrm{model}(\theta) - R_\textrm{crown}(\theta)\right)^2 d \theta,
\eeq
with $R_\textrm{model}$ and $R_\textrm{crown}$  the polar representations of the model and tree crown respectively. This procedure allows to find the best parameters $\alpha_g$ and $\alpha_p$. The same procedure is repeated with the additional constraint  $\alpha_p>0$.
  
This fitting method has been applied to 36 different tree crown pictures (see Supplementary Materials), out of which four representative cases have been extracted (figure \ref{comparison}).  
In figure \ref{comparison}a, the fit is fair with parametric values of  $\alpha_g$ and $\alpha_p$ both positive, as expected from what is known about shoot tropisms, namely upward gravitropism and positive phototropism. 
For some trees, the shape of the crown could not be well fitted with positive values of $\alpha_g$ and $\alpha_p$. However, allowing a negative value for  $\alpha_g$ (figure \ref{comparison}b) or $\alpha_p$ (figure \ref{comparison}c,d), leads to a good fit. 
Negative values for  $\alpha_p$ would mean a negative shoot phototropism, a condition that is usually not observed  in studies about shoot phototropism. We analyse in the discussion section the biological insights that such a situation may provide.
{Figure \ref{comparison}(c) and (d) provide two sets of parameters: the first one (red profile) for which we allowed any values for $\alpha_p$, the second one (black profile) for which $\alpha_p$ is constrained to be positive. These two comparisons clearly show that in some cases, like (c), the crown shape can be well fitted using several parameter sets (possibly even a continuum set of parameters), whereas others show a poor comparison with the real shape when $\alpha_p$ is constrained to be positive.} {On this matter, the diagonal described by equation \eqref{boundary2} and seen in figure \ref{diagram} plays a particular role: close to this line, shapes look similar, allowing for a family of solutions. }


\section{Discussion}


 Many tree species display characteristic and heritable crown shapes, at least at some stages and when trees have grown in isolation. Yet crown shape is remarkably variable with the environment (what biologists call "platicity"): trees grow differently if they are isolated in full sunlight or in a forest\cite{Cournede2008}, if they are submitted to wind or protected from it\cite{Davis1980}... Reconciling these two features has remained a challenge.  
Indeed there is a large number of  biological mechanisms that can  affect the intensity and orientation of shoot growth, e.g. phyllotaxis,  apical dominance \cite{edelin1996,Barthelemy2007}, response of buds break to light\cite{pierik2013}, changes in growth directions at different developmental stages, tropisms, internal correlations between organs\cite{Barthelemy2007}, as well as repulsive effects between branches mediated though phytochromes\cite{pierik2013}, biomechanical processes at the branch level, such as the bending of the branches under self-weight, its fixing through secondary growth and global re-orientating mechanisms related to reaction wood\cite{moulia2009}.
Moreover, the development of the crown is modifying the environment of the growing apices, creating a feedback of the actual crown shape over its development  (shading, wind sheltering...).  
All the challenge is then to identify the leading physical and physiological mechanisms which can explain the interaction between genetic control and sensitivity to the environment in the development of crown shape. Two standpoints can be adopted. 
  
One is to concentrate on the development of the branched architecture through bud breaks, shoot growth, and its remodelling through environmental factors interacting with the crown size and spread such as light competition or wind effects\cite{Mech1996,Eloy2017}. 
The second stand-point is to concentrate on what is occurring at the boundary of the crown, and more specifically on the apical buds that are sitting on its vicinity, and which are responsible for the increment in crown spread through the annual flush of shoot growth. This standpoint can be visualised when considering for example a coniferous tree in the springtime, on which the crown of the last year has kept its dark-green needles and the new sprouts undergoing primary growth (elongation) shows up with a much lighter green colour. The descriptive parameters of this crown increment are the distribution of the buds, the amount of their elongation, and their direction. And the interaction between the genetic specification and the local environment at the boundary of the crown is specified through the physiological processes that may influence these descriptive parameters.
  This second standpoint is the one that has been investigated in this work. Three different processes were thus retained:  the photomorphogenetic photosensitivity of shoot elongation (referred to as photosensitivity) and the two major tropisms known to affect the direction of the primary growth: photo- and gravi-tropism\cite{Bastien2013,Bastien2015}.  The initial orientation of the bud was considered to have negligible effect as growing shoot can reach their photo-gravitpropic set-point angle irrespective of the original orientation in an interval of time  that is very short compared to the morphogenesis of the crown shape (typically less than a day)\cite{Bastien2013,Bastien2015}.

In this article, we model this process of incremental primary growth at the boundary of the crown with a front propagation equation taking into account the effects of photosensitivity, phototropism and gravitropism, and the feedback of the actual crown shape through its shading effect of the growing apices, as a function of their position. We also included a synchronisation of the growth velocity among neighbouring shoots\cite{pierik2013,nagashima2012,moulia2015} resulting in a local flattening tendency of the canopy boundary, which is qualitatively similar to a surface tension. 
Through numerical simulations, we show that this front equation generally yields self-similar solutions at long time, independently of the initial shape. 
In the limit of vanishing surface tension, we show that these self-similar solutions can be described analytically by parametric curves, which, for particular values of the phototropic and gravitropic responses, can exhibit singularities. 

  
The outputs of the model were then compared to real tree crowns. As the dynamics of the crown growth is at the core of the model,  a direct comparison between the proposed front velocity, given by Eq.~\eqref{velocity}, and empirical observation of growth directions and amount would have been the best assessment. However, no such data was available, and their collection would require long term experiments. The assessment of the predictions of the model was thus achieved by comparing its self-similar solutions  with real tree crowns at the special stages at which the crown displays a typical and stationary shape (namely  the ``architectural unit stage'' and mature stages \cite{edelin1996,Barthelemy2007}). This was achieved on 36 crown images of different tree species sampling many taxonomic clades from different climates. Using the distance $d$ defined by Eq.~\eqref{eq:distance} measuring how the shapes extracted from the pictures differ from the analytical parametric curves, we could find, for each picture, the set of phototropic and gravitropic intensities that best fit the tree crown. It appears that in 97\% of the cases, we find a distance $d<0.05$, which we consider as good (see Supplementary Materials).


These results however raise two major questions: one about the values of the photo- and gravisensitivity parameters; and the second one about the significance of the two important assumptions of the model (aside from the choice of the three mechanisms): the linear photosensitivity curve and the uniform lighting.
{Regarding the photo- and gravisensitivity parameters, positive values of $\alpha_p$ and $\alpha_g$ allow for a satisfactory comparison with real crown shapes in only $33\%$ of the cases. For $75\%$ of the cases, a reasonable fit can be obtained with the constraint $\alpha_p\ge 0$. However negative values of $\alpha_p$  are  required to fit the remaining $25\%$  (figure \ref{comparison}d and Supplementary Materials).} Negative shoot phototropism  has never been documented in wild-type plants living in natural conditions and can be considered non-physiological. By the same token, negative shoot gravitropism is restrained to rare mutants, and is unlikely to be involved in the studied trees.  However, these negative values may reflect the influence of another tropism, namely auto-tropism\cite{moulia2009}.  Auto-tropism is the tendency for growing plant shoots to reduce their curvature, through a proprioceptive sensing\cite{Bastien2013,hamant2016}. It was found that the proprioceptive auto-tropism is a major player of the orientation of shoot growth\cite{Bastien2013,Bastien2015}. Auto-tropism was not included in our model and negative phototropism could compensate for that.
But it could also compensate for other processes acting at the whole branch level (e.g, branch biomechanical reorientation or shedding) that were not taken into account in the model. Some explicit representation of the growing shoot or of the whole branch would be required to asses if these additional mechanisms may  account for {the crown shapes that cannot be fitted with our model}. Therefore a new research avenue emerging from the present work is to extend the model to account for the average orientation of branches along the crown boundary, the effect of, for example,  auto-tropism, or self-weight \cite{moulia2009} as well as some mechanisms of spatial competition\cite{Runions2007} or of branch repulsion\cite{pierik2013}. In this work we have assessed  whether following the trajectories of surface points backward in time gives an idea of the underlying branches structure. But effects such as auto-tropism or bending under self-weight were not taken into account in our model and the ``inferred branch patterns'' were clearly not very realistic. 

The two other hypotheses of the model: i) linear photosensitivity of growth (the front velocity is assumed to be proportional to the average light intercepted) and ii) uniform lighting (light, on average, is assumed to be uniformly distributed in the upper hemisphere) remain also to be discussed. Indeed, whereas the  basis for the phototropic and gravitropic response lays on a wealth of published reports, as shown in the introduction section, the situation is less favourable here. Both hypotheses can be advocated to be a good starting point that allows to keep the model tractable and understandable (parsimony argument); yet they should not be {\it ad hoc}. 
The first hypothesis is  grounded in the fact that plants have light sensors called phytochromes that allow them to be sensitive to the ratio between the Red light and the Far Red light (R:FR) \cite{pierik2013}. FR light is almost fully reflected by foliage elements, whereas R light is largely absorbed. Then, for a finite front area $A$, the R:FR  ratio can be well approximated by the ratio of visible sky and of visible surrounding foliage. When $A \to 0$, assuming that the density of surrounding foliage is uniform, this ratio remains proportional to the angle of visible sky, and this corresponds to our model. In general it has been found that there is a linear relationship between the elongation rate of shoots and the R:FR ratio \cite{pierik2013}, which is usually negative for many open-habitat  species (leading to a shade avoiding) and positive for some tree species (a behavior known as shade tolerance)\cite{Gilbert2001}. 
The second  hypothesis seems much less straightforward. It is only realistic for fully overcast skies on cloudy days. On sunny days, the diffuse light from the sky displays a fairly uniform radiance,  but obviously the direct sunlight is highly non-uniform, with a bias towards the South (resp. North) in the Northern (resp. Southern) hemisphere \cite{Chelle2007}. And including a direct sunlight component in our model would clearly break the axi-symmetry of the crown shape due to a faster growth in the sun-facing side. On the contrary, tree crowns in Nature do not seem to be biased towards the sun \cite{Getzin2007}   (with the exception of {\it Araucaria columnaris}  \cite{Johns2017}). Is this a serious drawback for our model? Not necessarily. Indeed the phytochrome sensing of the R:FR ratio of the light is compensated for light intensity\cite{pierik2013}. Besides the elongation growth is known to be impaired by water stress, that is related to the direct sunlight, so that higher growth may actually happen on overcast cloudy or rainy days, and hence under uniform lighting\cite{Tardieu2003,Rinne1994}. Yet, to our knowledge, no explanation has ever been published to this puzzling question of crown axi-symmetry in isolated trees despite the asymmetry of many environmental factors (sunlight of course, but also evaporation, temperature...).  But now that we have shown that the two previous assumptions coupled with phototropism and gravitropism led to realistically looking tree crown shapes, it is clear that  investigating more thoroughly light-compensated sensing mechanisms and their relation to shoot growth and to crown shaping in trees is likely to be very rewarding. And including more mechanistic light responses in our model can be a tool to do so.

Despite this set of open points, our work clearly shows that the hypothesis that the observed crown shapes could be the result of a front propagation process driven by the light- and gravity- sensing at the crown boundary is a valid starting point. 
Moreover, it provides an insight on how to reconcile the genetic specificity and the environmentally-driven variability (plasticity) of crown shapes. 
Indeed if a tree has enough time to reach a stationary shape, then there should be no more influence of the initial state and/or the pathway to this form.  This stationary shape emerges from the recurrent effects of elongation photosensitivity and photo-and gravi-tropic sensitivities, and the feedback on these three processes of the interaction between the lighting environment and the actual crown shape. But the values of the three sensitivities are genetically-fixed. This may explain how a process driven by three responses to the environment (and hence allowing plasticity) yields a steady-state form that is heritable and probably genetically controlled. The situation is similar with what has been found for the shape of single shoots by\cite{Bastien2013,Bastien2015}. But for this to be achieved there should be no other player in the environment*crown shape interaction, a case only met when trees are grown in isolation and other factors --e.g. shading by neighbours, wind, frost-- are negligible. It is likely not to be the whole story though. Indeed there are indications that the steady shapes of the same tree at the (young) ``architectural unit stage'' and at the mature stage are different\cite{edelin1996,Barthelemy2007} (although this does not rely on observations of the same tree at the two stages, grown in isolation).  One may think that the shape control through photomorphogenetic and tropic mechanisms captured in our model only stabilises each of the two forms, whereas the shift between the two stages could be driven by changes in the branching development when trees developed their scaffold limbs through the process of developmental metamorphosis. To provide different attractors, though, non linear behaviour is required in our model; but there are many candidates for that (eg. interaction between different light sensors, phytochromes, cryptochromes\cite{pierik2013}\dots).

However, before that, our work also points at the need for data following the development of isolated trees from different species during decades. Only this type of data may allow to  assess whether self-similarity is reached. Additionally the effect of minor interventions, such as partial pruning, could be studied experimentally by looking at the shape dynamics and relaxation towards a self-similar shape.  Such detailed data are required to assess our model more accurately, as well as to assess the alternative models. It would be also very informative to have a full 3D shape (or at least view at right angles) to assess the axisymmetry of the crown and its robustness to environmental gradients across the crown.  This should be a priority now. But hopefully new tools such as terrestrial LIDARs or high-speed photogrametric devices are now available for that \cite{Popescu2003}. More detailed experimental studies of the diversity of photomorphogenetic and tropic responses for different species or under different environmental conditions would also be extremely useful (so that the sensitivities in our model could be measured independently rather then fitted ).

From a broader perspective, our approach also opens the way to interdisciplinary research to whole-plant morphogenesis. 
A first avenue is to change scale in our understanding of plant morphogenesis, form the scale of the meristem or single shoot  \cite{Bastien2013,hamant2016} to that of the whole tree. Advances on these aspects would help to further elucidate the mechanisms driving the natural variability of crown shapes \cite{Bastien2013,Chelakkot20170001} and how plants read their own shape\cite{hamant2016}.
Finally, in a different context, the approach proposed in the present article could be useful to study the growth of other biological systems that display crown growth, such as stony corals \cite{Kaandorp1996}, for which local growth velocities and direction depend both on internal regulation processes (nutrient exchange, for instance) and external cues (flow velocity, nutrient capture, etc\dots). 

\section*{Author contributions}
All authors conceived the study. L.D. and C.E. developed the numerical and analytical model. E.B. contributed to tree pictures. All authors wrote the paper. 

~

\section*{Acknowledgments}
We warmly thank C.~Coutand, A.~Lacointe, X.~Leoncini and B.~Roman for stimulating discussions.

\section*{Data accessibility}
The datasets analysed during the study are available from the corresponding author on reasonable request.

\section*{Funding statement}
The project leading to this publication has received funding from Excellence Initiative of Aix-Marseille University - A*MIDEX, a French ``Investissements d'Avenir'' programme. It has been carried out in the framework of the Labex MEC. 
We also acknowledge support from the CNRS (Mission pour l'interdisciplinarit\'e, project ARBRE).

\bibliography{all_ref}

\begin{thebibliography}{42}%
\makeatletter
\providecommand \@ifxundefined [1]{%
 \@ifx{#1\undefined}
}%
\providecommand \@ifnum [1]{%
 \ifnum #1\expandafter \@firstoftwo
 \else \expandafter \@secondoftwo
 \fi
}%
\providecommand \@ifx [1]{%
 \ifx #1\expandafter \@firstoftwo
 \else \expandafter \@secondoftwo
 \fi
}%
\providecommand \natexlab [1]{#1}%
\providecommand \enquote  [1]{``#1''}%
\providecommand \bibnamefont  [1]{#1}%
\providecommand \bibfnamefont [1]{#1}%
\providecommand \citenamefont [1]{#1}%
\providecommand \href@noop [0]{\@secondoftwo}%
\providecommand \href [0]{\begingroup \@sanitize@url \@href}%
\providecommand \@href[1]{\@@startlink{#1}\@@href}%
\providecommand \@@href[1]{\endgroup#1\@@endlink}%
\providecommand \@sanitize@url [0]{\catcode `\\12\catcode `\$12\catcode
  `\&12\catcode `\#12\catcode `\^12\catcode `\_12\catcode `\%12\relax}%
\providecommand \@@startlink[1]{}%
\providecommand \@@endlink[0]{}%
\providecommand \url  [0]{\begingroup\@sanitize@url \@url }%
\providecommand \@url [1]{\endgroup\@href {#1}{\urlprefix }}%
\providecommand \urlprefix  [0]{URL }%
\providecommand \Eprint [0]{\href }%
\providecommand \doibase [0]{http://dx.doi.org/}%
\providecommand \selectlanguage [0]{\@gobble}%
\providecommand \bibinfo  [0]{\@secondoftwo}%
\providecommand \bibfield  [0]{\@secondoftwo}%
\providecommand \translation [1]{[#1]}%
\providecommand \BibitemOpen [0]{}%
\providecommand \bibitemStop [0]{}%
\providecommand \bibitemNoStop [0]{.\EOS\space}%
\providecommand \EOS [0]{\spacefactor3000\relax}%
\providecommand \BibitemShut  [1]{\csname bibitem#1\endcsname}%
\let\auto@bib@innerbib\@empty
\bibitem [{\citenamefont {Barth{\'e}l{\'e}my}\ and\ \citenamefont
  {Caraglio}(2007)}]{Barthelemy2007}%
  \BibitemOpen
  \bibfield  {author} {\bibinfo {author} {\bibfnamefont {D.}~\bibnamefont
  {Barth{\'e}l{\'e}my}}\ and\ \bibinfo {author} {\bibfnamefont
  {Y.}~\bibnamefont {Caraglio}},\ }\href {\doibase 10.1093/aob/mcl260}
  {\bibfield  {journal} {\bibinfo  {journal} {Ann. Bot.}\ }\textbf {\bibinfo
  {volume} {99}},\ \bibinfo {pages} {375} (\bibinfo {year} {2007})}\BibitemShut
  {NoStop}%
\bibitem [{\citenamefont {Courn{\`e}de}\ \emph {et~al.}(2008)\citenamefont
  {Courn{\`e}de}, \citenamefont {Mathieu}, \citenamefont {Houllier},
  \citenamefont {Barth{\'e}l{\'e}my},\ and\ \citenamefont
  {de~Reffye}}]{Cournede2008}%
  \BibitemOpen
  \bibfield  {author} {\bibinfo {author} {\bibfnamefont {P.}~\bibnamefont
  {Courn{\`e}de}}, \bibinfo {author} {\bibfnamefont {A.}~\bibnamefont
  {Mathieu}}, \bibinfo {author} {\bibfnamefont {F.}~\bibnamefont {Houllier}},
  \bibinfo {author} {\bibfnamefont {D.}~\bibnamefont {Barth{\'e}l{\'e}my}}, \
  and\ \bibinfo {author} {\bibfnamefont {P.}~\bibnamefont {de~Reffye}},\ }\href
  {\doibase 10.1093/aob/mcm272} {\bibfield  {journal} {\bibinfo  {journal}
  {Ann. Bot.}\ }\textbf {\bibinfo {volume} {101}},\ \bibinfo {pages} {1207}
  (\bibinfo {year} {2008})}\BibitemShut {NoStop}%
\bibitem [{\citenamefont {Davis}(1980)}]{Davis1980}%
  \BibitemOpen
  \bibfield  {author} {\bibinfo {author} {\bibfnamefont {M.~L.}\ \bibnamefont
  {Davis}},\ }\href@noop {} {\bibfield  {journal} {\bibinfo  {journal} {Am.
  Midl. Nat.}\ ,\ \bibinfo {pages} {383}} (\bibinfo {year} {1980})}\BibitemShut
  {NoStop}%
\bibitem [{\citenamefont {M{\v{e}}ch}\ and\ \citenamefont
  {Prusinkiewicz}(1996)}]{Mech1996}%
  \BibitemOpen
  \bibfield  {author} {\bibinfo {author} {\bibfnamefont {R.}~\bibnamefont
  {M{\v{e}}ch}}\ and\ \bibinfo {author} {\bibfnamefont {P.}~\bibnamefont
  {Prusinkiewicz}},\ }\href@noop {} {\bibfield  {journal} {\bibinfo  {journal}
  {Proceedings of SIGGRAPH 96}\ ,\ \bibinfo {pages} {397}} (\bibinfo {year}
  {1996})}\BibitemShut {NoStop}%
\bibitem [{\citenamefont {Sierra-de Grado}\ \emph {et~al.}(1997)\citenamefont
  {Sierra-de Grado}, \citenamefont {Moulia}, \citenamefont {Fournier},
  \citenamefont {Al{\'\i}a},\ and\ \citenamefont
  {D{\'\i}ez-Barra}}]{sierra1997}%
  \BibitemOpen
  \bibfield  {author} {\bibinfo {author} {\bibfnamefont {R.}~\bibnamefont
  {Sierra-de Grado}}, \bibinfo {author} {\bibfnamefont {B.}~\bibnamefont
  {Moulia}}, \bibinfo {author} {\bibfnamefont {M.}~\bibnamefont {Fournier}},
  \bibinfo {author} {\bibfnamefont {R.}~\bibnamefont {Al{\'\i}a}}, \ and\
  \bibinfo {author} {\bibfnamefont {R.}~\bibnamefont {D{\'\i}ez-Barra}},\
  }\href@noop {} {\bibfield  {journal} {\bibinfo  {journal} {Trees-Structure
  and Function}\ }\textbf {\bibinfo {volume} {11}},\ \bibinfo {pages} {455}
  (\bibinfo {year} {1997})}\BibitemShut {NoStop}%
\bibitem [{\citenamefont {Bastien}\ \emph {et~al.}(2013)\citenamefont
  {Bastien}, \citenamefont {Bohr}, \citenamefont {Moulia},\ and\ \citenamefont
  {Douady}}]{Bastien2013}%
  \BibitemOpen
  \bibfield  {author} {\bibinfo {author} {\bibfnamefont {R.}~\bibnamefont
  {Bastien}}, \bibinfo {author} {\bibfnamefont {T.}~\bibnamefont {Bohr}},
  \bibinfo {author} {\bibfnamefont {B.}~\bibnamefont {Moulia}}, \ and\ \bibinfo
  {author} {\bibfnamefont {S.}~\bibnamefont {Douady}},\ }\href@noop {}
  {\bibfield  {journal} {\bibinfo  {journal} {Proc. Natl. Acad. Sci. U.S.A}\
  }\textbf {\bibinfo {volume} {110}},\ \bibinfo {pages} {755} (\bibinfo {year}
  {2013})}\BibitemShut {NoStop}%
\bibitem [{\citenamefont {Chauvet}\ \emph {et~al.}(2016)\citenamefont
  {Chauvet}, \citenamefont {Pouliquen}, \citenamefont {Forterre}, \citenamefont
  {Legu{\'e}},\ and\ \citenamefont {Moulia}}]{Chauvet2016}%
  \BibitemOpen
  \bibfield  {author} {\bibinfo {author} {\bibfnamefont {H.}~\bibnamefont
  {Chauvet}}, \bibinfo {author} {\bibfnamefont {O.}~\bibnamefont {Pouliquen}},
  \bibinfo {author} {\bibfnamefont {Y.}~\bibnamefont {Forterre}}, \bibinfo
  {author} {\bibfnamefont {V.}~\bibnamefont {Legu{\'e}}}, \ and\ \bibinfo
  {author} {\bibfnamefont {B.}~\bibnamefont {Moulia}},\ }\href@noop {}
  {\bibfield  {journal} {\bibinfo  {journal} {Scientific reports}\ }\textbf
  {\bibinfo {volume} {6}} (\bibinfo {year} {2016})}\BibitemShut {NoStop}%
\bibitem [{\citenamefont {Rivi{\`e}re}\ \emph {et~al.}(2017)\citenamefont
  {Rivi{\`e}re}, \citenamefont {Derr},\ and\ \citenamefont
  {Douady}}]{Riviere2017}%
  \BibitemOpen
  \bibfield  {author} {\bibinfo {author} {\bibfnamefont {M.}~\bibnamefont
  {Rivi{\`e}re}}, \bibinfo {author} {\bibfnamefont {J.}~\bibnamefont {Derr}}, \
  and\ \bibinfo {author} {\bibfnamefont {S.}~\bibnamefont {Douady}},\
  }\href@noop {} {\bibfield  {journal} {\bibinfo  {journal} {Phys. Biol.}\ }
  (\bibinfo {year} {2017})}\BibitemShut {NoStop}%
\bibitem [{\citenamefont {Chelakkot}\ and\ \citenamefont
  {Mahadevan}(2017)}]{Chelakkot20170001}%
  \BibitemOpen
  \bibfield  {author} {\bibinfo {author} {\bibfnamefont {R.}~\bibnamefont
  {Chelakkot}}\ and\ \bibinfo {author} {\bibfnamefont {L.}~\bibnamefont
  {Mahadevan}},\ }\href {\doibase 10.1098/rsif.2017.0001} {\bibfield  {journal}
  {\bibinfo  {journal} {J. Roy. Soc. . Interface}\ }\textbf {\bibinfo {volume}
  {14}} (\bibinfo {year} {2017}),\ 10.1098/rsif.2017.0001}\BibitemShut
  {NoStop}%
\bibitem [{\citenamefont {Goyal}\ \emph {et~al.}(2013)\citenamefont {Goyal},
  \citenamefont {Szarzynska},\ and\ \citenamefont {Fankhauser}}]{Goyal2013}%
  \BibitemOpen
  \bibfield  {author} {\bibinfo {author} {\bibfnamefont {A.}~\bibnamefont
  {Goyal}}, \bibinfo {author} {\bibfnamefont {B.}~\bibnamefont {Szarzynska}}, \
  and\ \bibinfo {author} {\bibfnamefont {C.}~\bibnamefont {Fankhauser}},\
  }\href@noop {} {\bibfield  {journal} {\bibinfo  {journal} {Trends Plant
  Sci.}\ }\textbf {\bibinfo {volume} {18}},\ \bibinfo {pages} {393} (\bibinfo
  {year} {2013})}\BibitemShut {NoStop}%
\bibitem [{\citenamefont {Perttunen}\ \emph {et~al.}(1998)\citenamefont
  {Perttunen}, \citenamefont {Siev{\"a}nen},\ and\ \citenamefont
  {Nikinmaa}}]{Perttunen1998}%
  \BibitemOpen
  \bibfield  {author} {\bibinfo {author} {\bibfnamefont {J.}~\bibnamefont
  {Perttunen}}, \bibinfo {author} {\bibfnamefont {R.}~\bibnamefont
  {Siev{\"a}nen}}, \ and\ \bibinfo {author} {\bibfnamefont {E.}~\bibnamefont
  {Nikinmaa}},\ }\href {\doibase 10.1016/S0304-3800(98)00028-3} {\bibfield
  {journal} {\bibinfo  {journal} {Ecol. Model.}\ }\textbf {\bibinfo {volume}
  {108}},\ \bibinfo {pages} {189} (\bibinfo {year} {1998})}\BibitemShut
  {NoStop}%
\bibitem [{\citenamefont {Allen}\ \emph {et~al.}(2005)\citenamefont {Allen},
  \citenamefont {Prusinkiewicz},\ and\ \citenamefont {DeJong}}]{Allen2005}%
  \BibitemOpen
  \bibfield  {author} {\bibinfo {author} {\bibfnamefont {M.~T.}\ \bibnamefont
  {Allen}}, \bibinfo {author} {\bibfnamefont {P.}~\bibnamefont
  {Prusinkiewicz}}, \ and\ \bibinfo {author} {\bibfnamefont {T.~M.}\
  \bibnamefont {DeJong}},\ }\href@noop {} {\bibfield  {journal} {\bibinfo
  {journal} {New Phytol.}\ }\textbf {\bibinfo {volume} {166}},\ \bibinfo
  {pages} {869} (\bibinfo {year} {2005})}\BibitemShut {NoStop}%
\bibitem [{\citenamefont {Barczi}\ \emph {et~al.}(2008)\citenamefont {Barczi},
  \citenamefont {Rey}, \citenamefont {Caraglio}, \citenamefont {De~Reffye},
  \citenamefont {Barthelemy}, \citenamefont {Dong},\ and\ \citenamefont
  {Fourcaud}}]{Barczi2008}%
  \BibitemOpen
  \bibfield  {author} {\bibinfo {author} {\bibfnamefont {J.}~\bibnamefont
  {Barczi}}, \bibinfo {author} {\bibfnamefont {H.}~\bibnamefont {Rey}},
  \bibinfo {author} {\bibfnamefont {Y.}~\bibnamefont {Caraglio}}, \bibinfo
  {author} {\bibfnamefont {P.}~\bibnamefont {De~Reffye}}, \bibinfo {author}
  {\bibfnamefont {D.}~\bibnamefont {Barthelemy}}, \bibinfo {author}
  {\bibfnamefont {Q.}~\bibnamefont {Dong}}, \ and\ \bibinfo {author}
  {\bibfnamefont {T.}~\bibnamefont {Fourcaud}},\ }\href@noop {} {\bibfield
  {journal} {\bibinfo  {journal} {Ann. Bot.}\ }\textbf {\bibinfo {volume}
  {101}},\ \bibinfo {pages} {1125} (\bibinfo {year} {2008})}\BibitemShut
  {NoStop}%
\bibitem [{\citenamefont {Palubicki}\ \emph {et~al.}(2009)\citenamefont
  {Palubicki}, \citenamefont {Horel}, \citenamefont {Longay}, \citenamefont
  {Runions}, \citenamefont {Lane}, \citenamefont {M{\v{e}}ch},\ and\
  \citenamefont {Prusinkiewicz}}]{Palubicki2009}%
  \BibitemOpen
  \bibfield  {author} {\bibinfo {author} {\bibfnamefont {W.}~\bibnamefont
  {Palubicki}}, \bibinfo {author} {\bibfnamefont {K.}~\bibnamefont {Horel}},
  \bibinfo {author} {\bibfnamefont {S.}~\bibnamefont {Longay}}, \bibinfo
  {author} {\bibfnamefont {A.}~\bibnamefont {Runions}}, \bibinfo {author}
  {\bibfnamefont {B.}~\bibnamefont {Lane}}, \bibinfo {author} {\bibfnamefont
  {R.}~\bibnamefont {M{\v{e}}ch}}, \ and\ \bibinfo {author} {\bibfnamefont
  {P.}~\bibnamefont {Prusinkiewicz}},\ }\href@noop {} {\bibfield  {journal}
  {\bibinfo  {journal} {ACM Trans. Graph.}\ }\textbf {\bibinfo {volume} {28}},\
  \bibinfo {pages} {58} (\bibinfo {year} {2009})}\BibitemShut {NoStop}%
\bibitem [{\citenamefont {Guo}\ \emph {et~al.}(2011)\citenamefont {Guo},
  \citenamefont {Fourcaud}, \citenamefont {Jaeger}, \citenamefont {Zhang},\
  and\ \citenamefont {Li}}]{Guo2011}%
  \BibitemOpen
  \bibfield  {author} {\bibinfo {author} {\bibfnamefont {Y.}~\bibnamefont
  {Guo}}, \bibinfo {author} {\bibfnamefont {T.}~\bibnamefont {Fourcaud}},
  \bibinfo {author} {\bibfnamefont {M.}~\bibnamefont {Jaeger}}, \bibinfo
  {author} {\bibfnamefont {X.}~\bibnamefont {Zhang}}, \ and\ \bibinfo {author}
  {\bibfnamefont {B.}~\bibnamefont {Li}},\ }\href@noop {} {\bibfield  {journal}
  {\bibinfo  {journal} {Ann. Bot.}\ }\textbf {\bibinfo {volume} {107}},\
  \bibinfo {pages} {723} (\bibinfo {year} {2011})}\BibitemShut {NoStop}%
\bibitem [{\citenamefont {Beyer}\ \emph {et~al.}(2014)\citenamefont {Beyer},
  \citenamefont {Letort},\ and\ \citenamefont {Courn{\`e}de}}]{Beyer2014}%
  \BibitemOpen
  \bibfield  {author} {\bibinfo {author} {\bibfnamefont {R.}~\bibnamefont
  {Beyer}}, \bibinfo {author} {\bibfnamefont {V.}~\bibnamefont {Letort}}, \
  and\ \bibinfo {author} {\bibfnamefont {P.-H.}\ \bibnamefont {Courn{\`e}de}},\
  }\href@noop {} {\bibfield  {journal} {\bibinfo  {journal} {Frontiers in plant
  science}\ }\textbf {\bibinfo {volume} {5}} (\bibinfo {year}
  {2014})}\BibitemShut {NoStop}%
\bibitem [{\citenamefont {Brower}\ \emph {et~al.}(1983)\citenamefont {Brower},
  \citenamefont {Kessler}, \citenamefont {Koplik},\ and\ \citenamefont
  {Levine}}]{brower1983geometrical}%
  \BibitemOpen
  \bibfield  {author} {\bibinfo {author} {\bibfnamefont {R.~C.}\ \bibnamefont
  {Brower}}, \bibinfo {author} {\bibfnamefont {D.~A.}\ \bibnamefont {Kessler}},
  \bibinfo {author} {\bibfnamefont {J.}~\bibnamefont {Koplik}}, \ and\ \bibinfo
  {author} {\bibfnamefont {H.}~\bibnamefont {Levine}},\ }\href@noop {}
  {\bibfield  {journal} {\bibinfo  {journal} {Phys. Rev. Lett.}\ }\textbf
  {\bibinfo {volume} {51}},\ \bibinfo {pages} {1111} (\bibinfo {year}
  {1983})}\BibitemShut {NoStop}%
\bibitem [{\citenamefont {Clavin}\ and\ \citenamefont
  {Searby}(2016)}]{clavin2016combustion}%
  \BibitemOpen
  \bibfield  {author} {\bibinfo {author} {\bibfnamefont {P.}~\bibnamefont
  {Clavin}}\ and\ \bibinfo {author} {\bibfnamefont {G.}~\bibnamefont
  {Searby}},\ }\href@noop {} {\emph {\bibinfo {title} {Combustion Waves and
  Fronts in Flows: Flames, Shocks, Detonations, Ablation Fronts and Explosion
  of Stars}}}\ (\bibinfo  {publisher} {Cambridge University Press},\ \bibinfo
  {year} {2016})\BibitemShut {NoStop}%
\bibitem [{\citenamefont {Pelc{\'e}}\ and\ \citenamefont
  {Libchaber}(2012)}]{Pelce2012a}%
  \BibitemOpen
  \bibfield  {author} {\bibinfo {author} {\bibfnamefont {P.}~\bibnamefont
  {Pelc{\'e}}}\ and\ \bibinfo {author} {\bibfnamefont {A.}~\bibnamefont
  {Libchaber}},\ }\href@noop {} {\emph {\bibinfo {title} {Dynamics of curved
  fronts}}}\ (\bibinfo  {publisher} {Elsevier},\ \bibinfo {year}
  {2012})\BibitemShut {NoStop}%
\bibitem [{\citenamefont {Edelin}\ \emph {et~al.}(1996)\citenamefont {Edelin},
  \citenamefont {Moulia},\ and\ \citenamefont {Tabourel}}]{edelin1996}%
  \BibitemOpen
  \bibfield  {author} {\bibinfo {author} {\bibfnamefont {C.}~\bibnamefont
  {Edelin}}, \bibinfo {author} {\bibfnamefont {B.}~\bibnamefont {Moulia}}, \
  and\ \bibinfo {author} {\bibfnamefont {F.}~\bibnamefont {Tabourel}},\
  }\href@noop {} {\bibfield  {journal} {\bibinfo  {journal} {Actes de
  l'{\'E}cole-Chercheur Inra en Bioclimatologie}\ }\textbf {\bibinfo {volume}
  {1}},\ \bibinfo {pages} {83} (\bibinfo {year} {1996})}\BibitemShut {NoStop}%
\bibitem [{\citenamefont {Bastien}\ \emph {et~al.}(2015)\citenamefont
  {Bastien}, \citenamefont {Douady},\ and\ \citenamefont
  {Moulia}}]{Bastien2015}%
  \BibitemOpen
  \bibfield  {author} {\bibinfo {author} {\bibfnamefont {R.}~\bibnamefont
  {Bastien}}, \bibinfo {author} {\bibfnamefont {S.}~\bibnamefont {Douady}}, \
  and\ \bibinfo {author} {\bibfnamefont {B.}~\bibnamefont {Moulia}},\
  }\href@noop {} {\bibfield  {journal} {\bibinfo  {journal} {Plos Comput.
  Biol.}\ }\textbf {\bibinfo {volume} {11}},\ \bibinfo {pages} {e1004037}
  (\bibinfo {year} {2015})}\BibitemShut {NoStop}%
\bibitem [{\citenamefont {Dumais}(2013)}]{Dumais2013}%
  \BibitemOpen
  \bibfield  {author} {\bibinfo {author} {\bibfnamefont {J.}~\bibnamefont
  {Dumais}},\ }\href {\doibase 10.1073/pnas.1219974110} {\bibfield  {journal}
  {\bibinfo  {journal} {Proc. Natl. Acad. Sci. U.S.A}\ }\textbf {\bibinfo
  {volume} {110}},\ \bibinfo {pages} {391} (\bibinfo {year}
  {2013})}\BibitemShut {NoStop}%
\bibitem [{\citenamefont {Pouliquen}\ \emph {et~al.}(2017)\citenamefont
  {Pouliquen}, \citenamefont {Forterre}, \citenamefont {B{\'e}rut},
  \citenamefont {Chauvet}, \citenamefont {Bizet}, \citenamefont {Legu{\'e}},\
  and\ \citenamefont {Moulia}}]{pouliquen2017}%
  \BibitemOpen
  \bibfield  {author} {\bibinfo {author} {\bibfnamefont {O.}~\bibnamefont
  {Pouliquen}}, \bibinfo {author} {\bibfnamefont {Y.}~\bibnamefont {Forterre}},
  \bibinfo {author} {\bibfnamefont {A.}~\bibnamefont {B{\'e}rut}}, \bibinfo
  {author} {\bibfnamefont {H.}~\bibnamefont {Chauvet}}, \bibinfo {author}
  {\bibfnamefont {F.}~\bibnamefont {Bizet}}, \bibinfo {author} {\bibfnamefont
  {V.}~\bibnamefont {Legu{\'e}}}, \ and\ \bibinfo {author} {\bibfnamefont
  {B.}~\bibnamefont {Moulia}},\ }\href@noop {} {\bibfield  {journal} {\bibinfo
  {journal} {Physical Biology}\ }\textbf {\bibinfo {volume} {14}},\ \bibinfo
  {pages} {035005} (\bibinfo {year} {2017})}\BibitemShut {NoStop}%
\bibitem [{\citenamefont {Pierik}\ and\ \citenamefont
  {de~Wit}(2013)}]{pierik2013}%
  \BibitemOpen
  \bibfield  {author} {\bibinfo {author} {\bibfnamefont {R.}~\bibnamefont
  {Pierik}}\ and\ \bibinfo {author} {\bibfnamefont {M.}~\bibnamefont
  {de~Wit}},\ }\href@noop {} {\bibfield  {journal} {\bibinfo  {journal}
  {Journal of Experimental Botany}\ }\textbf {\bibinfo {volume} {65}},\
  \bibinfo {pages} {2815} (\bibinfo {year} {2013})}\BibitemShut {NoStop}%
\bibitem [{\citenamefont {Moulia}\ \emph {et~al.}(2015)\citenamefont {Moulia},
  \citenamefont {Coutand},\ and\ \citenamefont {Julien}}]{moulia2015}%
  \BibitemOpen
  \bibfield  {author} {\bibinfo {author} {\bibfnamefont {B.}~\bibnamefont
  {Moulia}}, \bibinfo {author} {\bibfnamefont {C.}~\bibnamefont {Coutand}}, \
  and\ \bibinfo {author} {\bibfnamefont {J.-L.}\ \bibnamefont {Julien}},\
  }\href@noop {} {\bibfield  {journal} {\bibinfo  {journal} {Frontiers in plant
  science}\ }\textbf {\bibinfo {volume} {6}} (\bibinfo {year}
  {2015})}\BibitemShut {NoStop}%
\bibitem [{\citenamefont {Nagashima}\ and\ \citenamefont
  {Hikosaka}(2012)}]{nagashima2012}%
  \BibitemOpen
  \bibfield  {author} {\bibinfo {author} {\bibfnamefont {H.}~\bibnamefont
  {Nagashima}}\ and\ \bibinfo {author} {\bibfnamefont {K.}~\bibnamefont
  {Hikosaka}},\ }\href@noop {} {\bibfield  {journal} {\bibinfo  {journal} {New
  Phytologist}\ }\textbf {\bibinfo {volume} {195}},\ \bibinfo {pages} {803}
  (\bibinfo {year} {2012})}\BibitemShut {NoStop}%
\bibitem [{\citenamefont {Duchemin}\ and\ \citenamefont
  {Eggers}(2014)}]{Duchemin2014}%
  \BibitemOpen
  \bibfield  {author} {\bibinfo {author} {\bibfnamefont {L.}~\bibnamefont
  {Duchemin}}\ and\ \bibinfo {author} {\bibfnamefont {J.}~\bibnamefont
  {Eggers}},\ }\href@noop {} {\bibfield  {journal} {\bibinfo  {journal} {J.
  Comput. Phys.}\ }\textbf {\bibinfo {volume} {263}},\ \bibinfo {pages} {37 }
  (\bibinfo {year} {2014})}\BibitemShut {NoStop}%
\bibitem [{\citenamefont {Einstein}(2015)}]{Einstein2015}%
  \BibitemOpen
  \bibfield  {author} {\bibinfo {author} {\bibfnamefont {T.~L.}\ \bibnamefont
  {Einstein}},\ }\enquote {\bibinfo {title} {Handbook of crystal growth:
  Fundamentals},}\ \ (\bibinfo  {publisher} {Elsevier},\ \bibinfo {year}
  {2015})\ Chap.\ \bibinfo {chapter} {Equilibrium shape of crystals}, pp.\
  \bibinfo {pages} {215--264}\BibitemShut {NoStop}%
\bibitem [{\citenamefont {Pimpinelli}\ and\ \citenamefont
  {Villain}(1998)}]{Pimpinelli1998}%
  \BibitemOpen
  \bibfield  {author} {\bibinfo {author} {\bibfnamefont {A.}~\bibnamefont
  {Pimpinelli}}\ and\ \bibinfo {author} {\bibfnamefont {J.}~\bibnamefont
  {Villain}},\ }\href@noop {} {\emph {\bibinfo {title} {Physics of crystal
  growth}}},\ Vol.~\bibinfo {volume} {19}\ (\bibinfo  {publisher} {Cambridge
  University Press},\ \bibinfo {year} {1998})\BibitemShut {NoStop}%
\bibitem [{\citenamefont {Moulia}\ and\ \citenamefont
  {Fournier}(2009)}]{moulia2009}%
  \BibitemOpen
  \bibfield  {author} {\bibinfo {author} {\bibfnamefont {B.}~\bibnamefont
  {Moulia}}\ and\ \bibinfo {author} {\bibfnamefont {M.}~\bibnamefont
  {Fournier}},\ }\href@noop {} {\bibfield  {journal} {\bibinfo  {journal} {J.
  Exp. Bot.}\ }\textbf {\bibinfo {volume} {60}},\ \bibinfo {pages} {461}
  (\bibinfo {year} {2009})}\BibitemShut {NoStop}%
\bibitem [{\citenamefont {Moulia}\ \emph {et~al.}(1999)\citenamefont {Moulia},
  \citenamefont {Loup}, \citenamefont {Chartier}, \citenamefont {Allirand},\
  and\ \citenamefont {Edelin}}]{moulia1999}%
  \BibitemOpen
  \bibfield  {author} {\bibinfo {author} {\bibfnamefont {B.}~\bibnamefont
  {Moulia}}, \bibinfo {author} {\bibfnamefont {C.}~\bibnamefont {Loup}},
  \bibinfo {author} {\bibfnamefont {M.}~\bibnamefont {Chartier}}, \bibinfo
  {author} {\bibfnamefont {J.-M.}\ \bibnamefont {Allirand}}, \ and\ \bibinfo
  {author} {\bibfnamefont {C.}~\bibnamefont {Edelin}},\ }\href@noop {}
  {\bibfield  {journal} {\bibinfo  {journal} {Annals of Botany}\ }\textbf
  {\bibinfo {volume} {84}},\ \bibinfo {pages} {645} (\bibinfo {year}
  {1999})}\BibitemShut {NoStop}%
\bibitem [{\citenamefont {Eloy}\ \emph {et~al.}(2017)\citenamefont {Eloy},
  \citenamefont {Fournier}, \citenamefont {Lacointe},\ and\ \citenamefont
  {Moulia}}]{Eloy2017}%
  \BibitemOpen
  \bibfield  {author} {\bibinfo {author} {\bibfnamefont {C.}~\bibnamefont
  {Eloy}}, \bibinfo {author} {\bibfnamefont {M.}~\bibnamefont {Fournier}},
  \bibinfo {author} {\bibfnamefont {A.}~\bibnamefont {Lacointe}}, \ and\
  \bibinfo {author} {\bibfnamefont {B.}~\bibnamefont {Moulia}},\ }\href@noop {}
  {\bibfield  {journal} {\bibinfo  {journal} {Nature communications}\ }\textbf
  {\bibinfo {volume} {8}},\ \bibinfo {pages} {1014} (\bibinfo {year}
  {2017})}\BibitemShut {NoStop}%
\bibitem [{\citenamefont {Hamant}\ and\ \citenamefont
  {Moulia}(2016)}]{hamant2016}%
  \BibitemOpen
  \bibfield  {author} {\bibinfo {author} {\bibfnamefont {O.}~\bibnamefont
  {Hamant}}\ and\ \bibinfo {author} {\bibfnamefont {B.}~\bibnamefont
  {Moulia}},\ }\href@noop {} {\bibfield  {journal} {\bibinfo  {journal} {New
  Phytologist}\ }\textbf {\bibinfo {volume} {212}},\ \bibinfo {pages} {333}
  (\bibinfo {year} {2016})}\BibitemShut {NoStop}%
\bibitem [{\citenamefont {Runions}\ \emph {et~al.}(2007)\citenamefont
  {Runions}, \citenamefont {Lane},\ and\ \citenamefont
  {Prusinkiewicz}}]{Runions2007}%
  \BibitemOpen
  \bibfield  {author} {\bibinfo {author} {\bibfnamefont {A.}~\bibnamefont
  {Runions}}, \bibinfo {author} {\bibfnamefont {B.}~\bibnamefont {Lane}}, \
  and\ \bibinfo {author} {\bibfnamefont {P.}~\bibnamefont {Prusinkiewicz}},\
  }\href@noop {} {\bibfield  {journal} {\bibinfo  {journal} {NPH}\ }\textbf
  {\bibinfo {volume} {7}},\ \bibinfo {pages} {63} (\bibinfo {year}
  {2007})}\BibitemShut {NoStop}%
\bibitem [{\citenamefont {Gilbert}\ \emph {et~al.}(2001)\citenamefont
  {Gilbert}, \citenamefont {Jarvis},\ and\ \citenamefont
  {Smith}}]{Gilbert2001}%
  \BibitemOpen
  \bibfield  {author} {\bibinfo {author} {\bibfnamefont {I.~R.}\ \bibnamefont
  {Gilbert}}, \bibinfo {author} {\bibfnamefont {P.~G.}\ \bibnamefont {Jarvis}},
  \ and\ \bibinfo {author} {\bibfnamefont {H.}~\bibnamefont {Smith}},\
  }\href@noop {} {\bibfield  {journal} {\bibinfo  {journal} {Nature}\ }\textbf
  {\bibinfo {volume} {411}},\ \bibinfo {pages} {792} (\bibinfo {year}
  {2001})}\BibitemShut {NoStop}%
\bibitem [{\citenamefont {Chelle}\ and\ \citenamefont
  {Andrieu}(2007)}]{Chelle2007}%
  \BibitemOpen
  \bibfield  {author} {\bibinfo {author} {\bibfnamefont {M.}~\bibnamefont
  {Chelle}}\ and\ \bibinfo {author} {\bibfnamefont {B.}~\bibnamefont
  {Andrieu}},\ }\href@noop {} {\bibfield  {journal} {\bibinfo  {journal}
  {Frontis}\ ,\ \bibinfo {pages} {75}} (\bibinfo {year} {2007})}\BibitemShut
  {NoStop}%
\bibitem [{\citenamefont {Getzin}\ and\ \citenamefont
  {Wiegand}(2007)}]{Getzin2007}%
  \BibitemOpen
  \bibfield  {author} {\bibinfo {author} {\bibfnamefont {S.}~\bibnamefont
  {Getzin}}\ and\ \bibinfo {author} {\bibfnamefont {K.}~\bibnamefont
  {Wiegand}},\ }\href@noop {} {\bibfield  {journal} {\bibinfo  {journal}
  {Forest Ecology and Management}\ }\textbf {\bibinfo {volume} {242}},\
  \bibinfo {pages} {165} (\bibinfo {year} {2007})}\BibitemShut {NoStop}%
\bibitem [{\citenamefont {Johns}\ \emph {et~al.}(2017)\citenamefont {Johns},
  \citenamefont {Yost}, \citenamefont {Nicolle}, \citenamefont {Igic},\ and\
  \citenamefont {Ritter}}]{Johns2017}%
  \BibitemOpen
  \bibfield  {author} {\bibinfo {author} {\bibfnamefont {J.~W.}\ \bibnamefont
  {Johns}}, \bibinfo {author} {\bibfnamefont {J.~M.}\ \bibnamefont {Yost}},
  \bibinfo {author} {\bibfnamefont {D.}~\bibnamefont {Nicolle}}, \bibinfo
  {author} {\bibfnamefont {B.}~\bibnamefont {Igic}}, \ and\ \bibinfo {author}
  {\bibfnamefont {M.~K.}\ \bibnamefont {Ritter}},\ }\href@noop {} {\bibfield
  {journal} {\bibinfo  {journal} {Ecology}\ }\textbf {\bibinfo {volume} {98}},\
  \bibinfo {pages} {2482} (\bibinfo {year} {2017})}\BibitemShut {NoStop}%
\bibitem [{\citenamefont {Tardieu}(2003)}]{Tardieu2003}%
  \BibitemOpen
  \bibfield  {author} {\bibinfo {author} {\bibfnamefont {F.}~\bibnamefont
  {Tardieu}},\ }\href@noop {} {\bibfield  {journal} {\bibinfo  {journal}
  {Trends in plant Science}\ }\textbf {\bibinfo {volume} {8}},\ \bibinfo
  {pages} {9} (\bibinfo {year} {2003})}\BibitemShut {NoStop}%
\bibitem [{\citenamefont {Rinne}\ \emph {et~al.}(1994)\citenamefont {Rinne},
  \citenamefont {Saarelainen},\ and\ \citenamefont {Junttila}}]{Rinne1994}%
  \BibitemOpen
  \bibfield  {author} {\bibinfo {author} {\bibfnamefont {P.}~\bibnamefont
  {Rinne}}, \bibinfo {author} {\bibfnamefont {A.}~\bibnamefont {Saarelainen}},
  \ and\ \bibinfo {author} {\bibfnamefont {O.}~\bibnamefont {Junttila}},\
  }\href@noop {} {\bibfield  {journal} {\bibinfo  {journal} {Physiologia
  Plantarum}\ }\textbf {\bibinfo {volume} {90}},\ \bibinfo {pages} {451}
  (\bibinfo {year} {1994})}\BibitemShut {NoStop}%
\bibitem [{\citenamefont {Popescu}\ \emph {et~al.}(2003)\citenamefont
  {Popescu}, \citenamefont {Wynne},\ and\ \citenamefont
  {Nelson}}]{Popescu2003}%
  \BibitemOpen
  \bibfield  {author} {\bibinfo {author} {\bibfnamefont {S.~C.}\ \bibnamefont
  {Popescu}}, \bibinfo {author} {\bibfnamefont {R.~H.}\ \bibnamefont {Wynne}},
  \ and\ \bibinfo {author} {\bibfnamefont {R.~F.}\ \bibnamefont {Nelson}},\
  }\href@noop {} {\bibfield  {journal} {\bibinfo  {journal} {Canadian journal
  of remote sensing}\ }\textbf {\bibinfo {volume} {29}},\ \bibinfo {pages}
  {564} (\bibinfo {year} {2003})}\BibitemShut {NoStop}%
\bibitem [{\citenamefont {Kaandorp}\ \emph {et~al.}(1996)\citenamefont
  {Kaandorp}, \citenamefont {Lowe}, \citenamefont {Frenkel},\ and\
  \citenamefont {Sloot}}]{Kaandorp1996}%
  \BibitemOpen
  \bibfield  {author} {\bibinfo {author} {\bibfnamefont {J.~A.}\ \bibnamefont
  {Kaandorp}}, \bibinfo {author} {\bibfnamefont {C.~P.}\ \bibnamefont {Lowe}},
  \bibinfo {author} {\bibfnamefont {D.}~\bibnamefont {Frenkel}}, \ and\
  \bibinfo {author} {\bibfnamefont {P.~M.~A.}\ \bibnamefont {Sloot}},\ }\href
  {\doibase 10.1103/PhysRevLett.77.2328} {\bibfield  {journal} {\bibinfo
  {journal} {Phys. Rev. Lett.}\ }\textbf {\bibinfo {volume} {77}},\ \bibinfo
  {pages} {2328} (\bibinfo {year} {1996})}\BibitemShut {NoStop}%
\end{thebibliography}%

\end{document}